\documentclass{aa}  

\usepackage{natbib}
\bibpunct{(}{)}{;}{a}{}{,} 

\usepackage{array, multirow, graphicx}
\usepackage[dvipsnames]{xcolor}
\usepackage{wasysym}[integrals]
\usepackage{amssymb}
\usepackage{pifont}
\usepackage{siunitx}
\usepackage{booktabs, tabularx}
\usepackage{cellspace, makecell} 
\usepackage{mathtools}
\usepackage{txfonts}

\usepackage[]{hyperref}
\hypersetup{colorlinks,citecolor=blue, linkcolor=blue, urlcolor=blue, breaklinks=true}

\newcounter{magicrownumbers}

\newcommand{\lya}{Ly$\alpha$}

\newcommand{\hi}{\ion{H}{I}}
\newcommand{\civ}{\ion{C}{IV}}

\newcommand{\mgii}{\ion{Mg}{II}}
\newcommand{\abs}[1]{\left | #1 \right |}

\begin{document} 

   \title{MUSE Analysis of Gas around Galaxies (MAGG)}
   \subtitle{VII. Emission line galaxies near strong blended \lya\ absorption systems at $z\gtrsim3$}

    \titlerunning{Emission line galaxies near strong blended \lya\ absorption systems at $z\gtrsim3$}
    \authorrunning{M. Galbiati et al.}

   \author{Marta Galbiati\inst{1}\fnmsep\thanks{\email{marta.galbiati@unimib.it}}
          \and
          Davide Tornotti\inst{1}
          \and
          Michele Fumagalli\inst{1,2}
          \and
          Matteo Fossati\inst{1,3}
          \and
          Matthew Pieri\inst{4}
          }
          
\institute{Dipartimento di Fisica ``G. Occhialini'', Universit\`a degli Studi di Milano-Bicocca, Piazza della Scienza 3, I-20126 Milano, Italy
 \and
    INAF – Osservatorio Astronomico di Trieste, Via G. B. Tiepolo 11, I-34143 Trieste, Italy
\and
    INAF – Osservatorio Astronomico di Brera, Via Brera 28, I-21021 Milano, Italy
\and
    Aix Marseille Universite, CNRS, CNES, LAM, Marseille, France
}

   \date{\today}

\abstract{
We investigate the connection between strong, blended Ly$\alpha$ absorption systems (SBLAs) and $\approx1000$ Ly$\alpha$ emitting galaxies (LAEs) at $z\gtrsim3$ in 28 quasar fields from the MUSE Analysis of Gas around Galaxies (MAGG) survey. Selecting SBLAs as spectral regions with transmitted flux $-0.05<F<0.25$ over $\approx138\text{ km s}^{-1}$ bins, we find a strong correlation with LAEs within a projected distance of $R\le300\rm\,kpc$ and line-of-sight velocity separation of $|\Delta\varv|\le300 \text{ km s}^{-1}$. The association rate increases significantly with decreasing flux, a trend that persists also at smaller separations ($R<100$ kpc).  A two-dimensional cross-correlation analysis confirms significant clustering of LAEs around SBLAs, while no such clustering is seen for spectral regions with $F>0.25$. The correlation appears to also depend on the width of the spectral window used to identify SBLAs, with a larger window yielding a stronger signal. Our analysis confirms that SBLAs serve as probes of the CGM at the interface between the Ly$\alpha$ forest and the optically-thick Lyman limit systems. The significant dependence of the LAE-SBLA cross-correlation on the spectral binning used to select these absorbers motivates future tests of the current SBLA framework as a tracer of halos.}

   \maketitle
%

\section{Introduction}

The mapping of neutral hydrogen (\hi) using absorption spectroscopy of bright background quasars has led to a rich taxonomy of absorption line systems (ALSs), including high column densities Damped \lya\ Absorbers \citep[DLAs; $N_{\rm HI} \ge 10^{20.3}~\rm cm^{-2}$,][]{wolfe_damped_2005}; intermediate population of Super Lyman Limit Systems \citep[SLLSs also known as sub-DLAs; $N_{\rm HI} \ge 10^{19}~\rm cm^{-2}$,][]{peroux_nature_2002}; optically thick Lyman Limit Systems \citep[LLSs; $N_{\rm HI} \ge 10^{17.2}~\rm cm^{-2}$,][]{tytler_qso_1982}; and the optically thin \lya\ forest \citep{lynds_absorption-line_1971}. In addition to the empirical classification based on the observed column densities, the statistical properties of these various ALSs connect specific classes to identifiable cosmic structures. DLAs are often associated with the densest portion of the circumgalactic medium (CGM), broadly defined as the gas that extends beyond the galaxy disks and is confined within a few virial radii, including the disk-halo interface and the outer \hi\ disks of galaxies, especially at low redshifts \cite{wolfe_damped_2005,fumagalli_absorption-line_2011}. LLSs, with their partial ionization and inferred volume densities of $10^{-3}-10^{-1}~\rm cm^{-3}$ \citep{fumagalli_physical_2016}, are one of the best tracers of the $z\approx 3$ CGM but can also probe dense regions of the intergalactic medium (IGM) at higher redshift ($z\gtrsim 3.5$), where the mean neutral fraction of the Universe becomes significant enough to form slabs of partially neutral gas even at modest overdensities \citep{fumagalli_dissecting_2013}. Finally, the \lya\ forest is the tracer of the gas filaments that make up the IGM \citep{rauch_lyman_1998}.

However, nature does not make jumps, and the above scheme is bound to be incomplete or even inaccurate, especially at the transition points between classes. This is especially the case since the mapping between physical overdensities (i.e., the volume densities compared to the mean cosmological density) and column density is not a tight one-to-one relation due to local fluctuations arising from gas clumpiness or variation in the ionization field \citep{rahmati_evolution_2013}. The nature of the transition region occupied by SLLSs/sub-DLAs has been actively debated \citep{khare_nature_2007}. More recently, the transition between LLSs and the \lya\ forest has been the subject of renewed attention. The gas at \hi\ column densities of $10^{14}-10^{16}~\rm cm^{-2}$ is highly relevant for its contribution to the mean opacity of the Universe \citep{prochaska_definitive_2010}. Moreover, due to the multiphase nature of the CGM, which extends over a very wide range of densities and thermal and ionization conditions \citep{tumlinson_circumgalactic_2017}, the detailed contribution of halo gas in the transition region between LLSs and the \lya\ forest is becoming a relevant question.

This transition region has been the focus of a series of articles on the so-called ``strong, blended \lya\ forest absorption systems'' \citep[SBLAs,][]{Pieri2010, Pieri2014, Perez2023, Morrison2024}. This class of absorbers is identified via low transmission pixels at a set velocity scale, which can arise whenever multiple strong \lya\ lines are clustered in velocity. One of the suggested approaches to identify SBLAs is based on BOSS quasar spectra with a signal-to-noise ratio per pixel $S/N>3$ over a 100-pixel boxcar. Absorbed pixels with transmitted flux $F<0.25$, once rebinned by a factor of two into the SDSS velocity scale of $138~\rm km~s^{-1}$, select the SBLAs. Specific ranges of transmission further define subclasses of this population \citep[e.g., $-0.05<F<0.05$ representing the FS0 class of the strongest systems;][]{Pieri2014,Morrison2024}. DLAs are masked and excluded from the selection to avoid contamination. 
Due to their mixed nature, a combination of individual absorbers of different \hi\ and metal content contributes to the SBLA population, which can, however, be studied with spectral stacking \cite{Pieri2014}. Modeling of the Lyman limit of the SBLA composite spectrum suggests column densities $N_{\rm HI}<10^{16}-10^{16.5}~\rm cm^{-2}$ \cite{Pieri2014}. This limit can be refined further by considering the Lyman series lines, resulting in three different samples of SBLAs with increasing column density as a function of decreasing transmitted flux thresholds, $N_{\rm HI}\approx10^{15.11}$, $N_{\rm HI}\approx10^{15.64}$, and $N_{\rm HI}\approx10^{16.04}$ \citep{Morrison2024}. Thus, SBLAs are mainly optically thin systems that represent a transition population between LLSs and the weaker \lya\ forest. 

The analysis of the metal content of SBLAs provides further insight into the physical origin of this population. SBLAs are a heterogeneous class of metal absorbers composed of strong metal lines for $\approx 1/4$ of their population, especially in a low-ionization phase \citep{Morrison2024}. Higher ionization gas appears to be more widespread within SBLAs. Detailed photoionization modeling of SBLA composite spectra, under the assumption of optically thin gas, reveals a multiphase medium composed of at least three distinct phases of density and temperature: a low-ionization phase with $n_H = 1~\rm cm^{-3}$ and metallicity $[X/H]=0.8$, an intermediate-ionization phase with $n_H = 10^{-3}~\rm cm^{-3}$ and metallicity $[X/H]=-0.8$, and a high-ionization phase whose physical parameters are poorly constrained. 

Additional insight into the astrophysical structures that give rise to SBLAs originates from a clustering analysis \citep{Perez2023, Morrison2024}. The cross-correlation of SBLAs with the \lya\ forest provides a bias of $2.329\pm0.057$, consistent with that of DLAs. Thus, SBLAs are likely to arise from dark-matter overdensities comparable to halos of masses $\approx 0.5-1\times 10^{12}~\rm M_\odot$. This finding, together with evidence that SBLAs are a multiphase and enriched population, provides a strong link between this class of absorbers and the CGM of galaxies. Hence, SBLA and DLA systems probably trace different portions of the CGM of a similar population of galaxies \citep{Perez2023}. This hypothesis can be verified explicitly by cross-correlating SBLAs with galaxy populations. By performing this experiment in quasar fields with surveys of Lyman break galaxies (LBGs), \citet{Pieri2014} found that SBLAs with $F<0.25$ reside near an LBG on average 60 percent of the time. Therefore, SBLAs selected at $F< 0.25$ are ten times more likely to be associated with LBGs than spectral pixels with $F>0.75$, and are five times more likely to be near LBGs than for a random distribution. Moreover, in a recent simulation work using TNG50, \citet{santos_finding_2025} shows that SBLAs are highly efficient halo finders. In SDSS-like spectra, up to $\approx80\%$ of systems with low \lya\ transmission ($F<0.05$) reside within dark-matter halos, and they are associated with halos with masses of $\approx10^{12}\,M_\odot$. By adopting a hierarchical approach, they refined the selection of SBLAs, with larger systems consuming smaller ones, and obtained a narrower halo mass distribution compared to the non-hierarchical framework. This is consistent with the observational constraints derived by clustering analysis, reinforcing the interpretation of SBLAs as robust tracers of the CGM around massive galaxies. 

However, the statistics originating from the correlations of SBLAs and LBGs are significantly dependent on the completeness correction in the LBG surveys. Thanks to integral field spectrographs (IFSs) such as the Multi Unit Spectroscopic Explorer (MUSE) \citep{bacon_muse_2010} at the Very Large Telescope (VLT), dense and highly complete $z\gtrsim3$ redshift surveys are available. This enables further testing of the hypothesis that SBLAs trace the CGM of fairly massive cosmic structures. In this paper, we rely on the ``MUSE Analysis of Gas around Galaxies'' (MAGG) survey to test the link between \lya\ emitters (LAEs) and SBLAs in the redshift interval $2.5<z<4.5$. MAGG is a moderate depth MUSE/IFS redshift survey of 28 quasar fields at $z\approx 3.5-4$, for which high resolution and high $S/N$ quasar spectroscopy is available \citep{lofthouse_muse_2020}. At $z>3$, the MAGG survey has already revealed tight correlations between LAEs and strong \hi\, \civ, and a fraction of \mgii\ absorption line systems \citep{lofthouse_muse_2023,galbiati_muse_2023,galbiati_muse_2024}. However, the bulk of the ALSs arises not from the CGM of the LAEs, but from the large-scale gas filaments within which the galaxies are embedded. Moreover, MAGG uncovered a substantial dependence of the gas properties of LAEs on their galactic environment, with group galaxies embedded in more gas-rich regions (\citealt{lofthouse_muse_2023,galbiati_muse_2023,galbiati_muse_2024}; see also \citealt{dutta_muse_2020,dutta_metal-enriched_2021}).  

With the complete and homogeneous sample of X-Shooter spectra for the 28 MAGG quasars \citep{galbiati_muse_2024} and a catalog of $\approx 1,000$ LAEs with well-characterized completeness \citep{galbiati_muse_2023}, we investigate directly the link between SBLAs and moderate-mass, \lya\ emitting galaxies. The purpose of this study is, therefore, to provide additional information on whether this class of absorbers arises from the CGM, and which is the more likely population of parent halos. We structure this paper as follows. In Section~\ref{sec:data}, we present a brief overview of the available data. We describe the selection of SBLAs in the MAGG quasar spectra in Section~\ref{sec:sblas}. The results of the SBLA-LAE correlation are given in Section~\ref{sec:correlation}. A discussion of our findings and conclusions are provided in Section~\ref{sec:disc}. Throughout, unless otherwise noted, we quote magnitudes in the AB system and distances in physical units. We adopt $\Omega_{\mathrm{m}}=0.307$ and $H_0=67.7\rm\,km\,s^{-1}\,Mpc^{-1}$ \citep{Planck2016}. 

\section{Spectroscopic observations}
\label{sec:data}

\subsection{Quasar absorption spectroscopy}\label{sec:data_spectra}

The MAGG survey combines VLT/MUSE observations between $465-930$~nm at resolution $R\approx 2000-3500$ of 28 fields centered on $z\approx3.2-4.5$ quasars with high-resolution ($ R\gtrsim30,000$) and high or moderate signal-to-noise spectroscopy ($S/N\gtrsim10$ per pixel) of the central quasars. The full dataset as well as the detailed steps of the data reduction are described by \cite{lofthouse_muse_2020}. In this work, we make use of the VLT/X-shooter \citep{Vernet2011} spectra of the MAGG quasars observed in the UVB ($300-559.5$~nm and $R=7450$ with $1\rlap{.}{\arcsec}0$ slit) and VIS ($559.5-1024$~nm and $R=4350$ with $0\rlap{.}{\arcsec}9$ slit) arms. Archival spectroscopy from the data release XQ-100 survey is available for 13 of the 28 quasars \citep{Lopez2016} and on the ESO archive for two additional sightlines (J$015741-010629$ and J$020944+051713$). The X-Shooter observations of the remaining quasars (PID 0109.A$-$0559; PI M. Galbiati) are described in \citet{galbiati_muse_2024}. The details, such as the wavelength coverage, the final S/N ratio, and the spectral resolution, are listed in table 2 in \citet{lofthouse_muse_2020}, and in table 1 in \citet{galbiati_muse_2024}. All the information on the observations and data reduction of the archival spectra can be found in \citet{Lopez2016}, while \citet{galbiati_muse_2024} details the reduction of the newer observations. 

    \begin{figure}
    	\centering
   	\resizebox{.9\hsize}{!}{
	\includegraphics{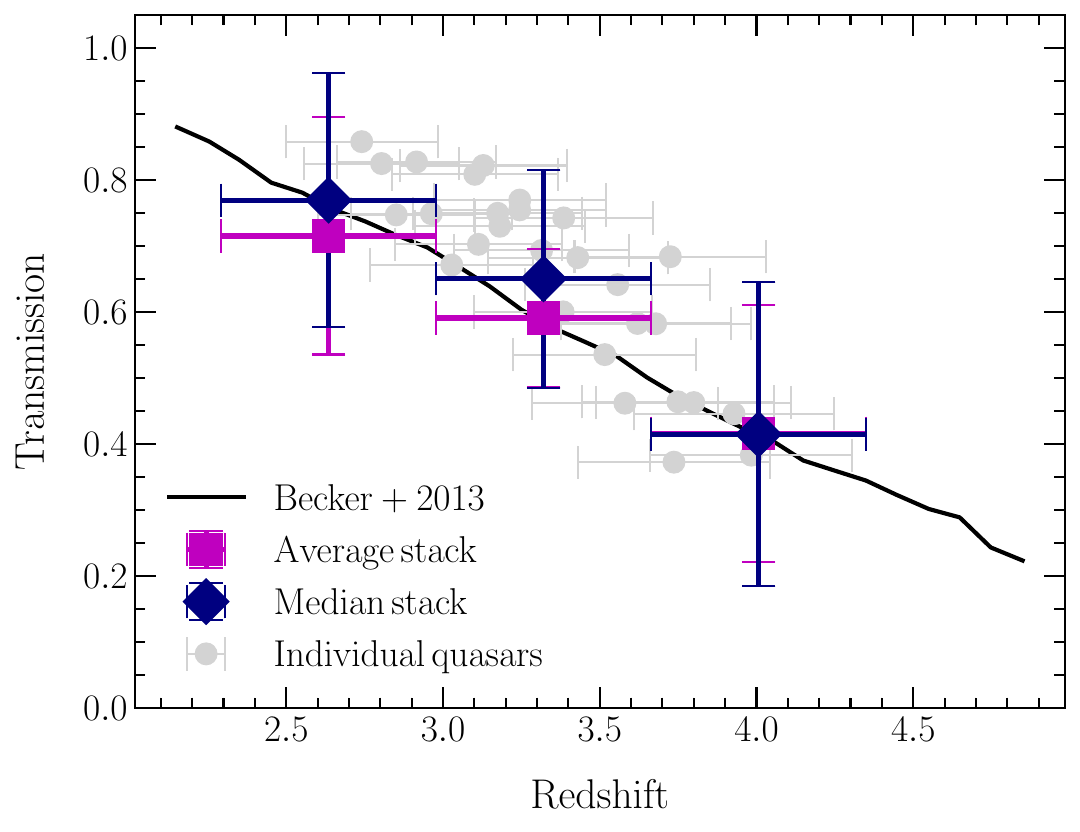}}
       \caption{Average transmitted flux in the rest-frame region $1041~\AA~<\lambda_{\rm RF,QSO}<1185$~\AA\ of the quasar spectra, as a function of the \lya\ redshift, compared to the average cosmic value measured by \citet{Becker2013}. Shown are the mean (purple) and median (violet) stacked spectra in three \lya\ redshift intervals within $2.5\lesssim z\lesssim4.5$. The uncertainties reproduce the standard deviation in each bin. In the background,  there is the average transmission of individual spectra (in grey).}
       \label{fig:transmission}
   \end{figure}

As the selection of SBLAs relies on the relative flux transmission, we require an accurate placement of the quasar continuum. To model the quasar continuum and normalize the spectra, knots have been manually selected in regions of the quasar continuum free of absorption and used for a cubic spline fit. The details of the continuum modeling and the associated validation tests are presented in \citet{lofthouse_muse_2020} and \citet{ galbiati_muse_2024}. In the present work, we further verified that the mean flux transmission in the  \lya\ forest is in agreement with the average values reported by \citet{Becker2013}. We measured the average transmission within the quasar rest-frame wavelength range $1041<\lambda_{\rm RF,QSO}<1185$~\AA\ in both the individual and stacked $3.0\lesssim z_{\rm QSO}\lesssim4.5$ quasar spectra (see Fig. \ref{fig:transmission}). We observe that the average transmission of the stacked spectra agrees with that of \citet{Becker2013} within $1\sigma$ and only find a marginal offset, of the order of $\approx13$~percent, in the individual ones. We verified that the results of this work do not change significantly if such a correction factor is applied to the spectra; hence, we regard the selection of SBLAs robust with respect to the continuum placement. 

\subsection{Catalogue of \lya\ emitters}
\label{sec:data_LAEs}

In the MAGG survey, \citet{galbiati_muse_2023} identified 921 \lya\ emitting galaxies in the redshift range $2.8<z<6.6$ (median $z\approx3.91$). For each galaxy, we measured its \lya\ luminosity (median $\log[L_{\rm Ly\alpha}/({\rm erg\,s^{-1}})]\approx42.10$) and projected distance from the central quasar of each field (median $R\approx160\rm\,kpc$). In particular, the fluxes are estimated using the curve of growth analysis in pseudo-narrow band images. The map from \citet{Schlafly2011} and the extinction law from \citet{Fitzpatrick1999} are used to correct them for the Milky Way extinction. We also derived an estimate of the redshift from their \lya\ emission lines and took that of the red peak in case of double-peaked line profiles. These galaxies are typically located at the low-mass end of the galaxies' mass function at these redshifts, with $M_\star=10^9-10^{10}{\rm \, M_\odot}$, and are hosted in dark matter halos of $M_{\rm h}=10^{10}-10^{11.5}{\rm \, M_\odot}$ \citep[see, e.g.,][]{Ouchi2020,herrero_alonso_clustering_2023, Herrera2025}.

We selected these galaxies following two key steps \citep[see,][for the details]{galbiati_muse_2023}. First, we run {\sc sextractor} \citep{Bertin1996} on the MUSE white-light image to identify the sources emitting bright rest-frame UV continuum. The output segmentation map was then used to extract the spectra. By using M. Fossati's fork\footnote{\url{https://matteofox.github.io/Marz}} of {\sc marz} \citep{Hinton2016}, we assigned a reliable spectroscopic redshift to 1200 sources. Second, we searched for \lya\ emitting galaxies. We subtracted all these continuum-detected sources from the MUSE datacubes and used the {\sc CubExtractor} package \citep{Cantalupo2019} to identify emission lines of at least 27 connected voxels with signal-to-noise ratio $S/N\ge3$. We included in the final sample all the sources detected at integrated signal-to-noise ratio $ISN\ge7$ and with line profiles that are not consistent with \ion{C}{IV}, \ion{C}{III}] or [\ion{O}{II}] emitters (e.g., in case of double peaks) or other lower-redshift transitions.

\section{Selection of SBLAs}
\label{sec:sblas}

    \begin{figure}
    	\centering
   	\resizebox{\hsize}{!}{
	\includegraphics{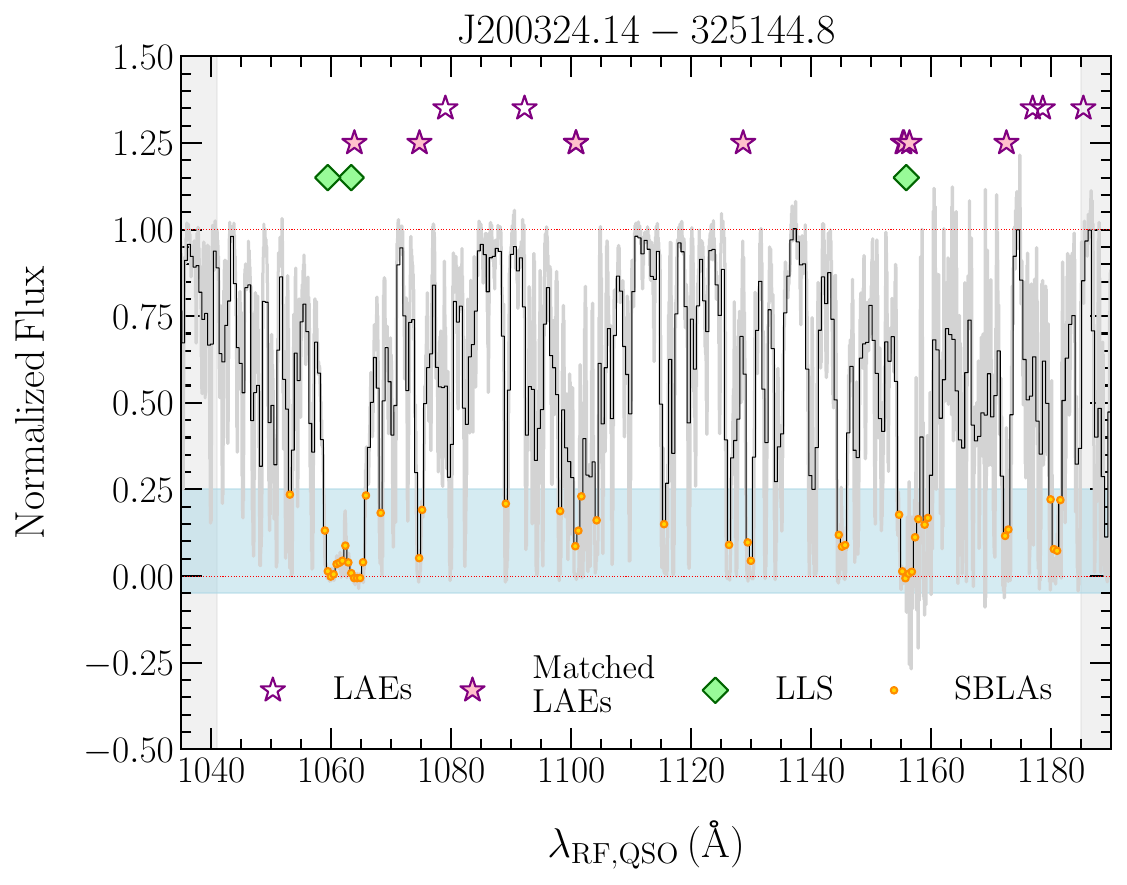}}
       \caption{Results of the search for SBLAs in the X-Shooter spectrum of the quasar J200324.14-325144.8. The original spectrum (lightgray) has been continuum-normalized, convolved with a Gaussian filter to match the resolution of BOSS, and re-sampled into a wavelength grid with a constant bin size of $138~\rm km\,s^{-1}$ (black). The pixels selected as SBLAs (orange) in the region $1041~\AA<\lambda_{\rm RF,QSO}<1185$~\AA\ (that is, excluding the gray shaded regions) have a transmitted flux within the range $-0.05<F<0.25$ (blue shaded region). Known LLS are shown as green diamonds. We also show LAEs that are matched (filled purple stars) and not matched (empty purple stars) to at least one SBLA within $R<300\rm\,kpc$ and $\lvert\Delta\varv\rvert\le300\rm\,km\,s^{-1}$ in Section~\ref{sec:correlation}.}
       \label{fig:SBLA_example}
   \end{figure}

In this work, we select SBLAs as \lya\ forest systems with transmission $-0.05<F<0.25$ over regions of $\approx138\rm\,km\,s^{-1}$ \citep[see][]{Pieri2014}, arising from strong and blended \lya\ absorption lines. Such systems have been previously studied in SDSS-III/BOSS \citep{Pieri2014} and SDSS-IV/eBOSS spectra \citep{Perez2023, Morrison2024}, for which the size of a spectral resolution element is ${\rm FWHM}\approx138\rm\,km\,s^{-1}$ (approximately twice the standard SDSS spectral pixel). The original definition was calibrated at the BOSS resolution, with the adopted velocity window matching a single instrumental resolution element. This choice mitigates pixel-scale noise fluctuations while preventing multiple identifications within the same resolution element. Similarly, the adopted transmission range was empirically defined for BOSS-like noise properties, ensuring that saturated pixels affected by noise are not artificially excluded. This behaviour is illustrated in the simulated spectra presented in \citet{Pieri2014}. In order to enable a direct comparison with these previous works, we apply the same selection criteria to the continuum-normalized X-Shooter spectra. We then discuss how the higher spectral resolution of X-Shooter may influence the resulting SBLA sample. As a first step, since both the UVB and VIS arms have higher resolving power (see, Section~\ref{sec:data_spectra}) than BOSS spectrographs ($1650\lesssim R\lesssim2150$), we convolve the X-Shooter spectra with a Gaussian filter to match the resolution of BOSS. We then re-sample the convolved spectra to a wavelength grid with constant log-spaced bins with pixel size of $138\rm\,km\,s^{-1}$ by conserving the total flux \citep[see, e.g.,][]{Carnall2017}. We chose this grid as it matches that of BOSS spectra rebinned by a factor of $\times2$, as done by \citet{Pieri2014} to improve the $S/N$ of the absorber selection. We then used the catalog of DLAs, with their respective column densities, detected in MAGG by \citet{lofthouse_muse_2023} to mask these strong \lya\ absorption systems. The same work also provides a catalog of LLSs, which we do not mask, but will be used to investigate their degree of overlap in the final sample of SBLAs and compare it with the predictions from \citet{Pieri2014}.

We restrict our search of SBLAs to the region $1041~\AA<\lambda_{\rm RF,QSO}<1185$~\AA\ in the rest-frame of the quasars as it is free from contamination of any high order Lyman series line other than \lya\, and it is not significantly affected by continuum fitting noise by excluding Ly$\beta$ and \ion{O}{VI} emission lines and the quasar proximity zone \citep[see also][]{Morrison2024}. Finally, we selected SBLAs as systems with transmitted flux $-0.05<F<0.25$ over wavelength bins of $138\rm\,km\,s^{-1}$ (see Fig. \ref{fig:SBLA_example} for an example). Finally, in defining the range of \lya\ transmission used to select SBLAs, we note an important caveat. As shown in Fig.~\ref{fig:transmission}, the mean IGM transmission decreases by a factor of $\approx2$ from $z=3$ to $z=4$, which means that a fixed flux threshold for selecting SBLAs may include systems that satisfy the criteria due to the stronger IGM absorption at higher redshift rather than because of their intrinsic properties. A redshift-dependent transmission threshold should be calibrated in future works to address this limitation.
\begin{figure}
\centering
\resizebox{\hsize}{!}{
\includegraphics{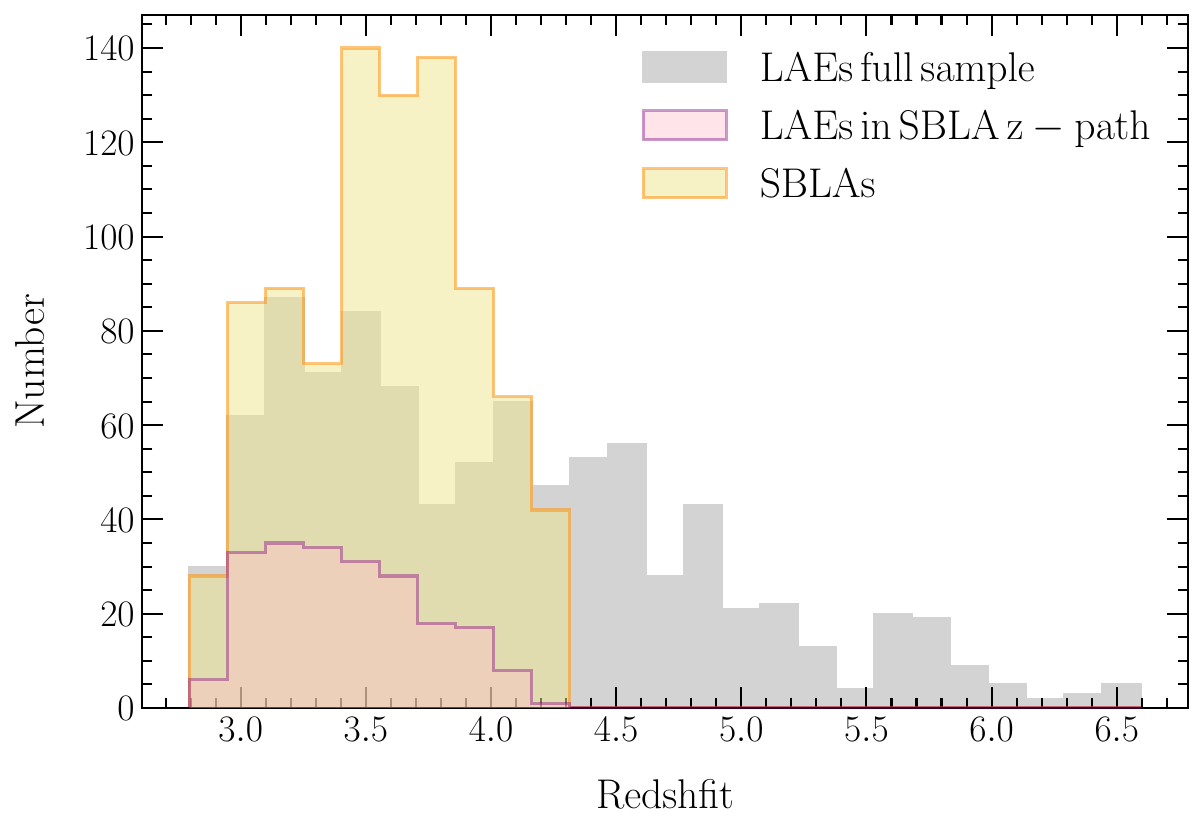}}
\caption{ Redshift distribution of the full LAE sample (grey), the subset of LAEs selected along the SBLA path (purple), and the complete SBLA sample (orange).}
\label{fig:histo-laes-sblas}
\end{figure}
A summary of the resulting samples is shown in Fig.~\ref{fig:histo-laes-sblas}, where we present the redshift distributions of the full LAE sample, the LAEs selected along the redshift path of the SBLAs, and the SBLA sample.

\section{The SBLA-LAE correlation}
\label{sec:correlation}

\subsection{Fractions of SBLAs associated with LAEs}
 
In \citet{Pieri2014}, associations with galaxies are defined within a projected distance $R\le300\rm~kpc$ and line-of-sight separation $\abs{\Delta\varv}\le300\rm~km~s^{-1}$, and for consistency we adopt the same definition. We note, however, that the MUSE field of view extends for $\approx 250$~kpc in radius, which implies that most of the LAEs detected are in fact considered when searching for associations. The region between $\approx 250-300$~kpc is covered only by the four corners of the field of view, and hence it is more poorly sampled. We also note that the expected virial radius of LAEs with masses up to $\approx 10^{11}~M_\odot$ \citep[see, e.g.,][]{herrero_alonso_clustering_2023, Herrera2025} is $\approx 35~\rm kpc$ at $z\approx3$, hence the linking length we adopted are likely probing regions that extend far beyond the CGM (the thresholds correspond to $\gtrsim8$ times the virial radius, and $\approx3$ times the virial velocity, respectively). Fig.~\ref{fig:SBLA_example} offers an example of SBLA-LAE associations in a MAGG field. 

     \begin{figure*}
    	\centering
   	\resizebox{\hsize}{!}{
	\includegraphics{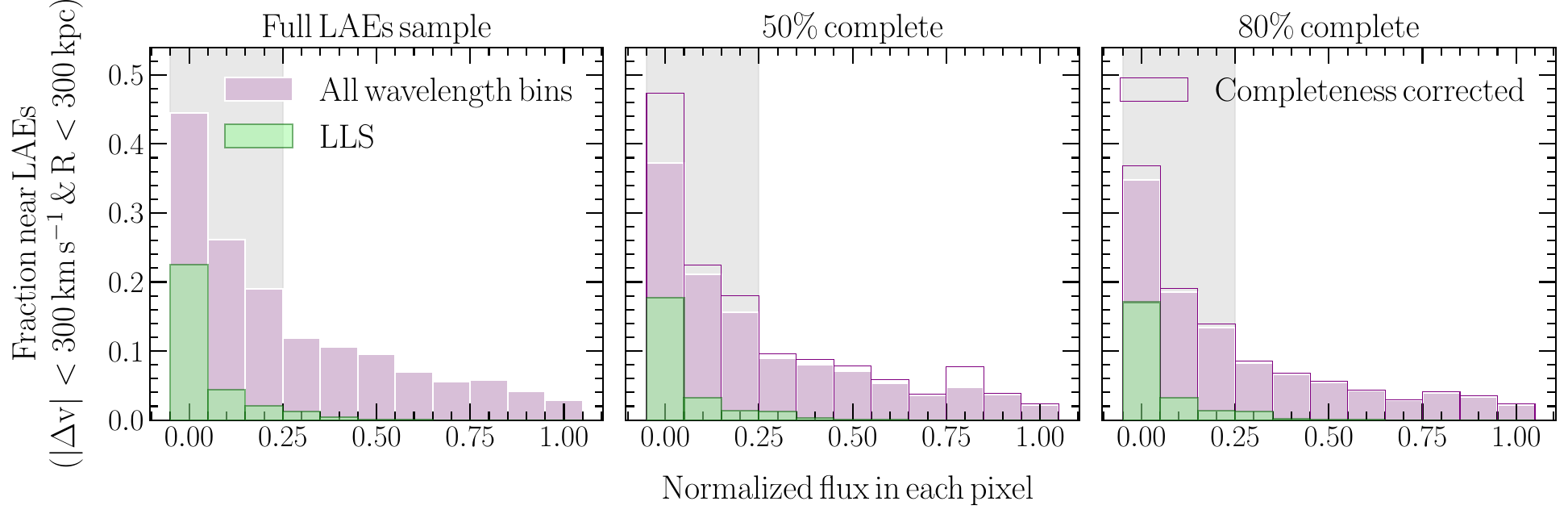}}
       \caption{Fraction of $138~\rm km\,s^{-1}$ wavelength bins, as a function of their transmitted fluxes, that are associated with at least one LAE within a projected distance $R\le300\rm\,kpc$ and line-of-sight separation $\abs{\Delta\varv}\le300\rm km\,s^{-1}$. The pink histograms show the results for all the absorption systems, while the green ones show the contribution to each flux interval of the wavelength bins that are also identified as LLSs. The gray shaded area highlights the wavelength bins with fluxes $-0.05<F<0.25$ that are identified as SBLAs. In the three panels, we show associations with the full sample of MAGG LAEs (left panel), and those above the $50$~percent (middle panel) and $80$~percent (right panel) completeness limit. Fractions computed with the respective completeness corrections are also shown (purple histogram).}
       \label{fig:correlation}
   \end{figure*}

\begin{table}[!t]
\def\arraystretch{1.15}
\caption{Statistics of LAEs and wavelength bins with $F<0.25$ and $\Delta v_{\rm spec}=138\rm\,km\,s^{-1}$ associated with each others within $\abs{\Delta\varv}\le300\rm\,km\,s^{-1}$ and $R\le300\rm\,kpc$.}  
\label{tbl:association}    
\centering
\resizebox{1\hsize}{!}{
\begin{tabular}{l c c c}
\toprule
\toprule
& $-0.05<F<0.05$ & $0.05<F<0.15$ & $0.15<F<0.25$ \\
\midrule
$N_{\rm SBLAs}^{(a)}$ & 164 & 275 & 442 \\
$f_{\rm SBLAs}^{(b)}$ & 0.45 & 0.27 & 0.19 \\
\bottomrule
\end{tabular}
}
\tablefoot{$^{(a)}$Total number of wavelength bins per interval of transmission. $^{(b)}$Fraction of wavelength bins identified in each interval of transmission that is also associated with at least one LAE.}
\end{table}

We show in Fig.~\ref{fig:correlation} (pink histogram in the left panel) and report in Table~\ref{tbl:association} the fraction of $138\rm\,km\,s^{-1}$ wide wavelength bins, as a function of their transmitted flux, that is found to be associated with at least one LAE within the limits above. The fractions of absorption systems identified as SBLAs ($-0.05<F<0.25$) connected to LAEs rises from $\approx20$~percent for wavelength bins with fluxes $0.15<F<0.25$, to $\approx25$~percent for bins with $0.05<F<0.15$ and reaches $\approx45$~percent for bins with $-0.05<F<0.05$. Instead, only $\lesssim10$~percent of the wavelength bins with $F\gtrsim0.25$ are associated with galaxies; that is, these bins are over three times less likely to be found near LAEs. Conversely, 211 LAEs are found in the redshift path of the wavelength bins, and $\approx57$~percent is associated with at least one $-0.05<F<0.25$ SBLA.

We next test the robustness of our results against the completeness of the sample of LAEs. We employed the selection function from \citet{fossati_muse_2021} and \citet{galbiati_muse_2023}, which gives the completeness as a function of both the \lya\ luminosity and the redshift of the galaxies ((meaning our ability to recover LAEs in MUSE observations). To this end, we compute the fraction of LAEs associate to different wavelength bins considering only those above the $50$~percent and $80$~percent completeness limit, which corresponds to luminosities of $\log[L_{\rm Ly\alpha}/({\rm erg\,s^{-1}})]=41.87$ and $\log[L_{\rm Ly\alpha}/({\rm erg\,s^{-1}})]=42$ at $z=3.5$, respectively. The results are displayed as pink histograms in the middle and right panels of Fig.~\ref{fig:correlation}. We then apply a correction by rescaling the number of LAEs found to be associated with SBLAs according to the completeness of the sample at the \lya\ luminosity of the galaxies (purple histograms in Fig.~\ref{fig:correlation}). We observe that the completeness of the sample of LAEs does not significantly impact the results. Indeed, the fraction of SBLAs with fluxes $-0.05<F<0.05$ ($0.05<F<0.15$) that are associated with galaxies corresponds to $\gtrsim30$~percent ($\gtrsim20$~percent) even when requiring the sample of LAEs to be $50$~percent complete.

     \begin{figure}
    	\centering
   	\resizebox{\hsize}{!}{
	\includegraphics{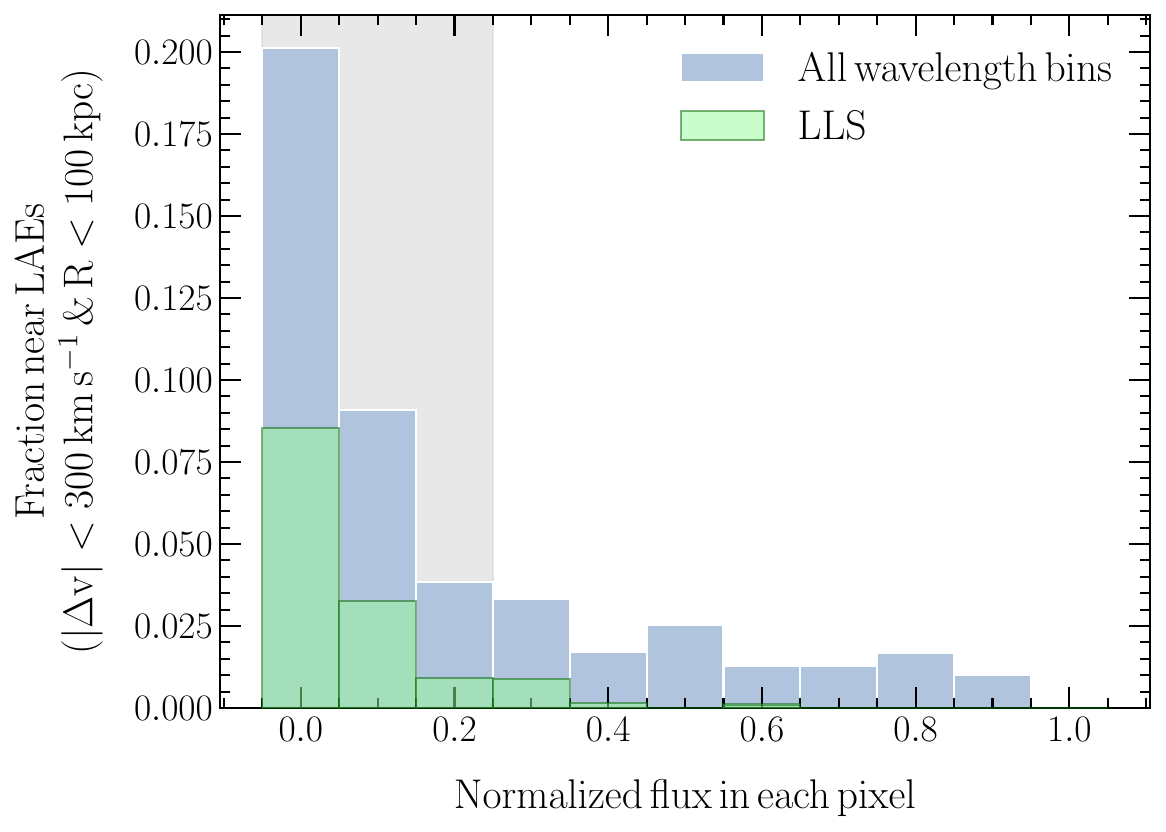}}
       \caption{Same as the left panel in Fig.~\ref{fig:correlation}, but for LAEs detected within 100~kpc of the quasar line of sight.}
       \label{fig:correlation_100kpc}
   \end{figure}

  \begin{figure*}
    	\centering
	\includegraphics[width=.8\textwidth]{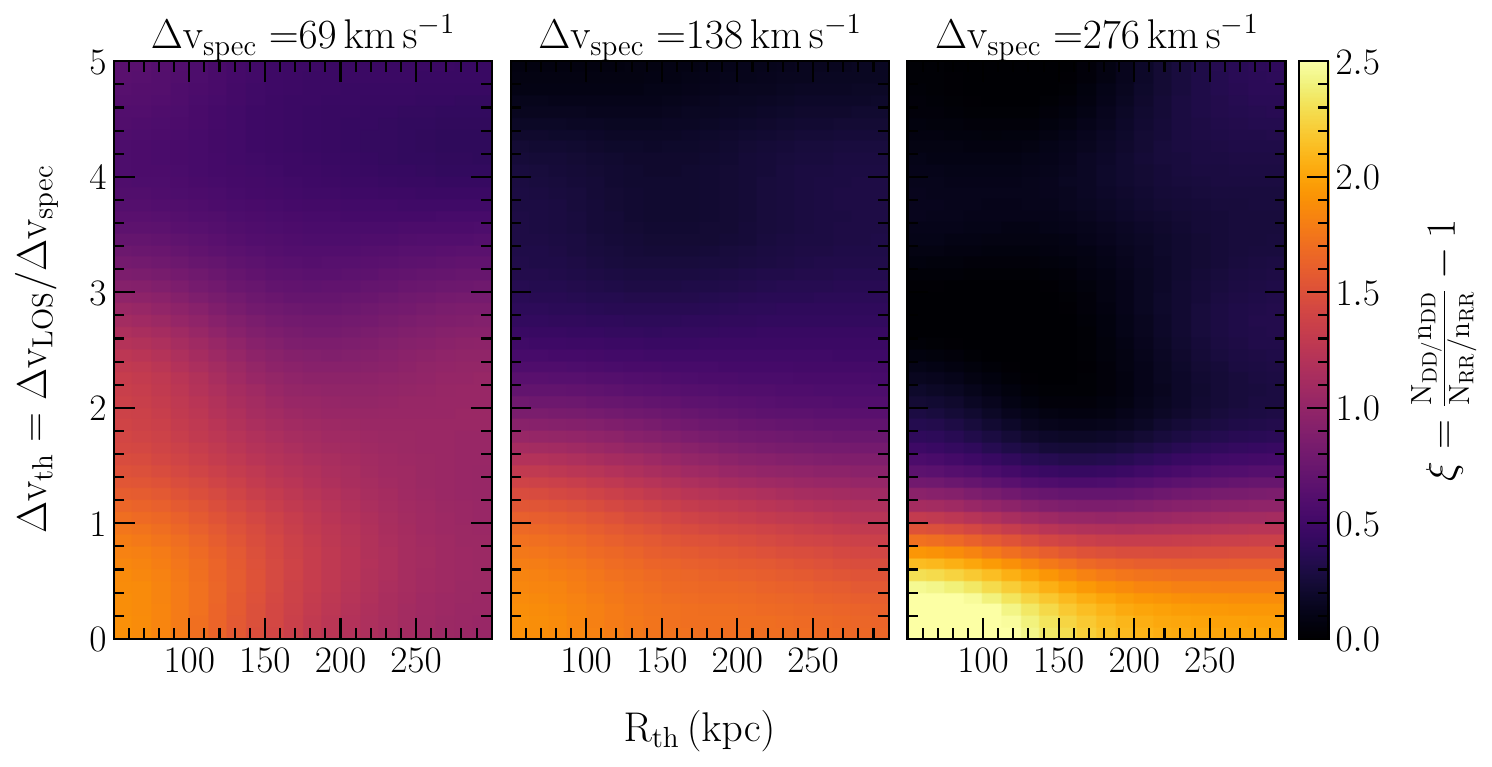} \\
    \includegraphics[width=.8\textwidth]{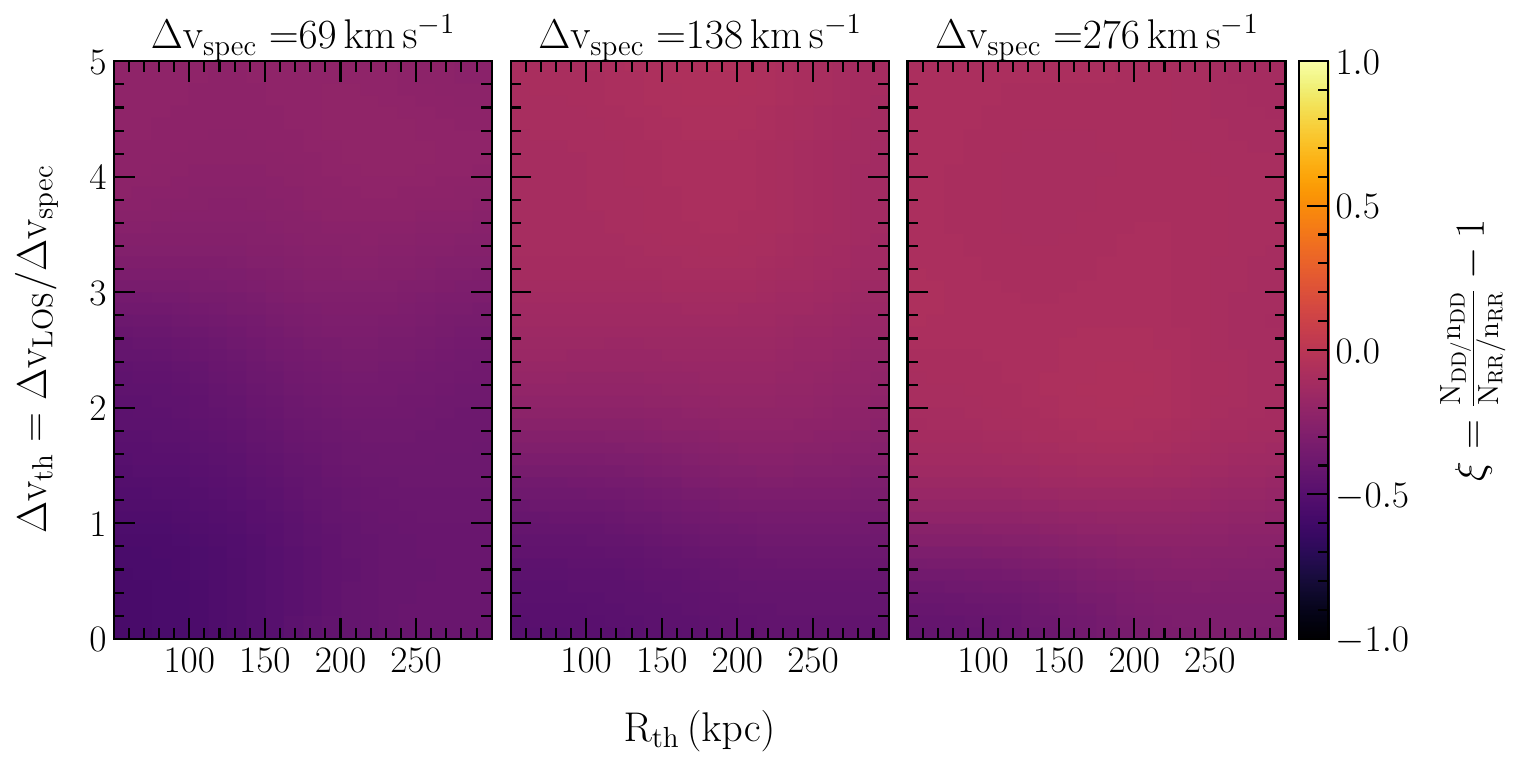}
       \caption{Two-dimensional LAE-absorber cross-correlation functions, $\xi$. {\it Upper panels:} Cross-correlation functions for systems with $-0.05<F<0.25$. The linking velocity window, $\Delta\varv_{\rm th}$, and projected separation, $\Delta R_{\rm th}$, vary along the vertical and horizontal axis, respectively. Each of the three panels shows the results for SBLAs identified in quasar spectra with a pixel size of $\Delta v_{\rm spec}=69\rm\,km\,s^{-1}$ (left panel), $138\rm\,km\,s^{-1}$ (middle panel), and $276\rm\,km\,s^{-1}$ (right panel). Line of sight separations $\Delta\varv_{\rm th}<\Delta v_{\rm spec}$ are masked. {\it Lower panels:} Cross-correlation functions between LAEs and systems with $F>0.25$. Note the different limits on the $\xi$ color scale.}
       \label{fig:correlation_links}
   \end{figure*}

To assess the contribution of SBLAs that arises from gas at smaller separations from LAEs, i.e., in the CGM on scales $\lesssim 3R_{\rm vir}$, we repeat the analysis above but selecting only associations within $\approx 100~\rm kpc$. As shown in Fig.~\ref{fig:correlation_100kpc}, the trend observed in Fig.~\ref{fig:correlation} persists also for associations at a smaller impact parameter. Wavelength bins with $-0.05<F<0.05$ are more often associated with galaxies than spectral regions at higher transmission. This indicates a tendency to find more absorbed bins in the CGM of galaxies, although the overall fraction of SBLAs associated with LAEs decreases by a factor of $\approx 2$ compared to the analysis at $R\leq 300~\rm kpc$. 

By means of the catalog of strong \hi\ absorbers identified in the MAGG survey by \citet{lofthouse_muse_2023}, we can next investigate the contribution of LLSs to the results we obtained. Indeed, while DLAs have already been masked, a fraction of wavelength bins with low transmitted flux and identified as SBLAs can actually be strong \hi\ absorbers, such as LLS. To quantify the degree of overlap, we derive the contribution of LLSs to the SBLAs-LAEs association by computing the fraction of wavelength bins connected to LAEs that are also LLSs, as a function of the transmitted flux (green histograms in Fig.~\ref{fig:correlation}). We found that out of the $45$~percent of wavelength bins with $-0.05<F<0.05$ associated with LAEs, $\approx20$~percent of them are also LLSs. As expected, such contribution decreases steeply for increasing transmitted flux, down to $\lesssim2$~percent for bins with $F>0.05$. As also noted before, this result does not vary significantly with the completeness of the LAEs sample, nor with the impact parameter used to establish the associations. 

Overall, we estimate the contribution of LLSs to be of the order of $\approx17.3$~percent in the wavelength bins with transmitted fluxes $-0.05<F<0.25$ that are identified as SBLAs. This small fraction confirms that systems at lower column density than LLSs drive the bulk of the observed associations, and hence we are not merely recovering the clustering of LAEs around optically-thick absorbers \citep{lofthouse_muse_2023}. Conversely, all LLSs are SBLAs, as expected by the marked \lya\ absorption core of these optically thick absorbers. Our analysis reaffirms the conclusion of previous authors that put the incidence of LLSs in the SBLAs population below $10~$percent. \citet{Pieri2014} estimates that around $\approx3.7$~percent of their blended absorbers sample is optically thick, even when assuming that $100$~percent of the LLSs are selected as SBLAs. 
Within MAGG, where we have a complete census of LLS, we confirm this result. Once we account for the correction in the incidence of LLSs discussed in \citet{lofthouse_muse_2023} as a result of the MAGG preselection, we find that the contribution of LLSs to the sample of SBLAs is $\approx4.8$~percent.

\subsection{The 2D cross-correlation functions}
\label{ssec:xcorr}

To investigate up to what degree the cross-match between SBLAs and LAEs depends on the re-binning of the spectra, we reproduce the steps of Section~\ref{sec:sblas} rebinning the X-Shooter spectra at their native resolution in pixels that have a width corresponding to half (that is, $69\rm\,km\,s^{-1}$, i.e. $\approx 1.7$ times UVB X-shooter resolution) and twice ($276\rm\,km\,s^{-1}$) the size of those defined by \citet{Pieri2014}. We also explore how varying the limits in projected and velocity space within which we connect the absorption systems to the galaxies affects the SBLA-LAE association rate. 

We compute the two-dimensional SBLA-LAE projected cross-correlation function ($\xi$) using the \citet{Davis1983} estimator\footnote{In this formalism, the cross-correlation is defined as $\xi=\frac{N_{\rm DD}/n_{\rm DD}}{N_{\rm RR}/n_{\rm RR}}-1$, where $N_{\rm DD}$, and $N_{\rm RR}$ are the number of galaxy-absorber pairs identified in the data and in the random sample, respectively, within each separation interval, while $n_{\rm DD}$, and $n_{\rm RR}$ account for the total number of pairs in the two samples.}. This represents the excess probability of finding an LAE within line-of-sight separation $\Delta\varv_{\rm th}$ and distance $R_{\rm th}$ from an SBLA, compared to the probability of finding an LAE within the same volume in a random place of the Universe. To reproduce the expectations for a sample of randomly distributed SBLAs and LAEs, we connected the absorbers identified in the spectrum of a given quasar with LAEs detected in a field randomly extracted among those centered on the other quasars \citep[as done in previous MAGG papers, e.g.,][]{galbiati_muse_2023}. We then bootstrapped 1,000 times over the fields and took the median values of the resulting distribution. The cross-correlation functions that we obtained with this method are shown in Fig.\ref{fig:correlation_links}. 

In the upper panels, we computed the cross-correlation by searching for SBLAs ($-0.05<F<0.25$) in the quasar spectra re-binned in the three different pixel sizes introduced above. In the lower panel, we show the cross-correlations obtained by linking to LAEs the pixels with $F>0.25$. To ensure a consistent comparison, the line-of-sight velocity separation in all panels is computed as a multiple of the spectral pixel size. Values at impact parameters $\gtrsim 250~\rm kpc$ should be treated with caution, as we are approaching the edge of the MUSE field of view and thus have a sparser sample.  

 \begin{figure*}
    	\centering
	\includegraphics[width=.54\textwidth]{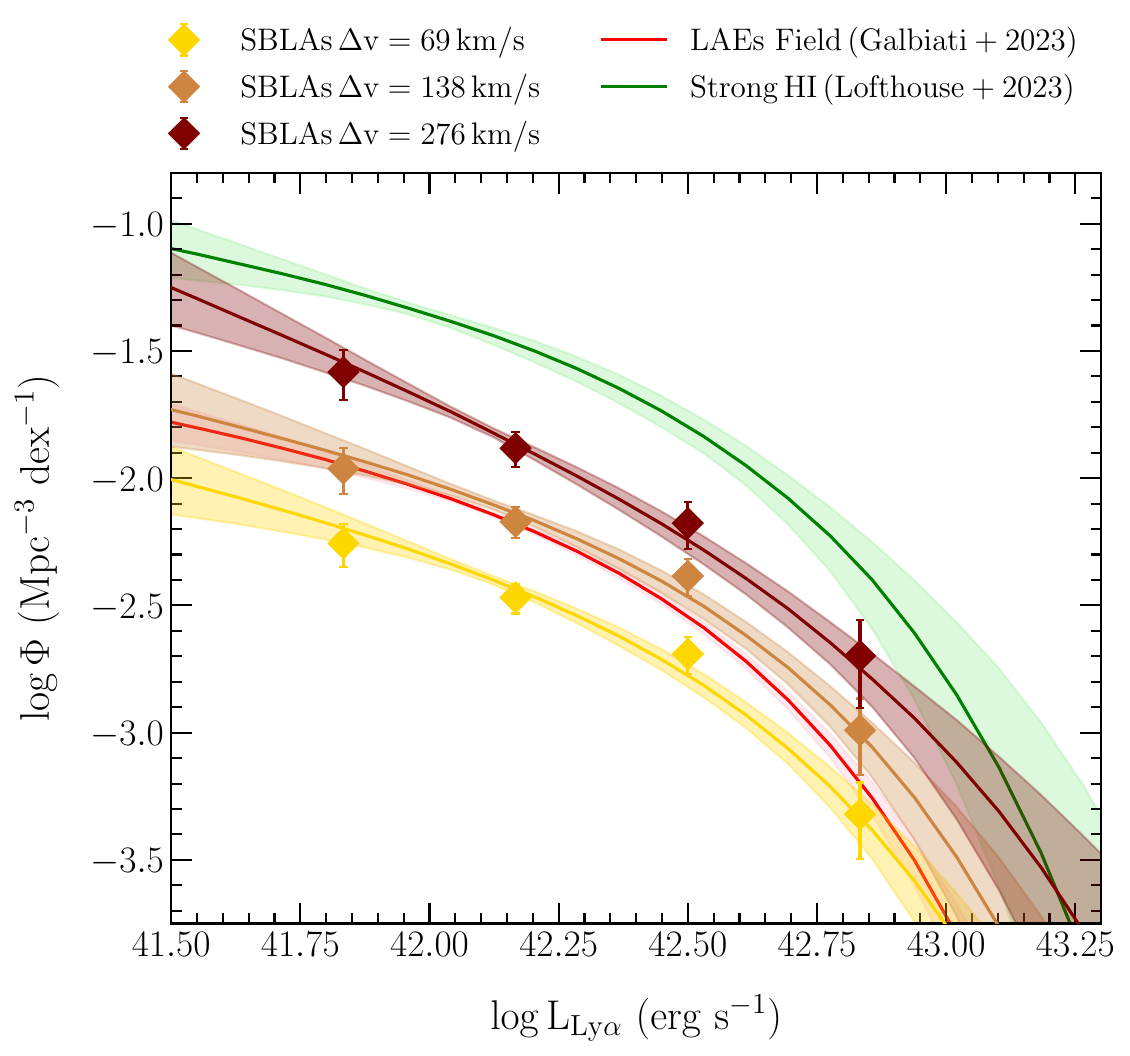}
    \includegraphics[width=.45\textwidth]{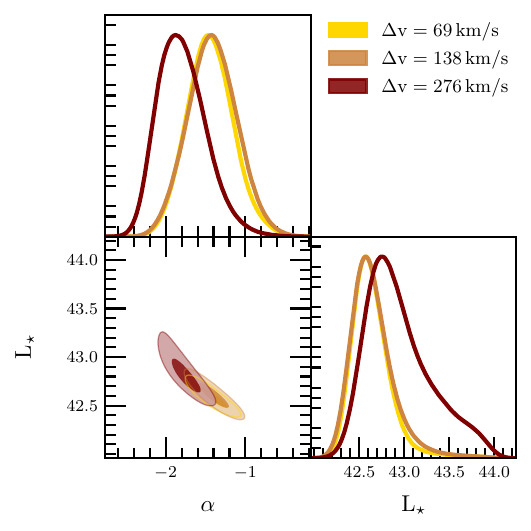}
       \caption{\lya\ luminosity functions of the LAEs associated with SBLAs within $R\le300\rm\,kpc$ and $\abs{\Delta\varv}\le300\rm~km~s^{-1}$. {\it Left panel}: \lya\ luminosity functions of the LAEs associated with SBLAs identified in the quasar spectra binned to $\Delta v_{\rm spec}=69,\,138,\,276\rm\,km\,s^{-1}$ (from light to dark colors). For comparison, we also show the luminosity function expected for LAEs in the field \citep[red,][]{galbiati_muse_2023} and for those found around strong \hi\ absorbers \citep[green,][]{lofthouse_muse_2023}. {\it Right panel}: posterior distribution of the $\alpha$ and $L_\star$ free parameters obtained fitting the luminosities functions with a \citet{Schechter1976}. The contours mark the confidence intervals corresponding to $1\sigma$ and $2\sigma$.}
       \label{fig:LF}
   \end{figure*}

Examining Fig.~\ref{fig:correlation_links}, we notice that LAEs are more likely to be found near SBLAs compared to random regions, with a clear preference to cluster towards low velocities and small impact parameters. Conversely, there is no obvious clustering of LAEs in spectral bins with $F>0.25$, i.e., near non-SBLAs. The distribution of LAEs around these spectral bins is consistent at any separation with $\xi\approx0$, that is, with a random distribution. It is thus more likely that systems with $F>0.25$ arise from IGM gas. Both the amplitude and the radial profile of the clustering of LAEs around SBLAs vary with the size of the spectral pixels. The excess of LAEs is larger around SBLAs identified in spectra with increasing $\Delta v_{\rm spec}$ and reaches its maximum around $\Delta v_{\rm spec}=276\rm\,km\,s^{-1}$. We explicitly tested that further increasing the spectral pixel size results in a roughly constant cross-correlation amplitude (with variations of $\lesssim 15\%$). Although the SBLA fraction at $\Delta v_{\rm spec}=276\rm\,km\,s^{-1}$ has dropped by $\approx50\%$ compared to that at $69\rm\,km\,s^{-1}$, such observed cross-correlation trend is driven by the survival of SBLAs that are made of smaller ones selected at lower $\Delta v_{\rm spec}$. In other words, at large spectral pixel sizes, the observed cross-correlation amplitude reflects the persistence of more extended structures, rather than the appearance of new ones. Finally, we note that the scale over which we detect a significant LAE-SBLA correlation, $R\lesssim 150\,\rm kpc$, extends beyond the typical virial radius expected for halos hosting LAEs ($\approx35\rm\,kpc$). This implies that part of the signal may arise from the CGM of neighboring clustered galaxies rather than only the halo of an individual LAE itself.

\subsection{The LAE luminosity function near SBLAs}

A complementary approach to studying the projected correlation function to connect LAEs and SBLAs is examining the LAE luminosity function in regions that host SBLAs. Once again, we consider SBLAs as a function of the size of the spectral pixels and statistically study how LAEs distribute in their surroundings as a function of the \lya\ luminosity. Following the steps detailed in \citet{fossati_muse_2021} and \citet{galbiati_muse_2023}, we compute the \lya\ luminosity functions of LAEs identified within $\abs{\Delta\varv}\le300\rm\,km\,s^{-1}$ and $R\le300\rm\,kpc$ around SBLAs and fit them with a parametric \citet{Schechter1976} function. Results are shown in Fig.~\ref{fig:LF} and reported in Table~\ref{tab:LF_results}. These are compared with the luminosity functions of LAEs representative of the field from \citet{galbiati_muse_2023}, as well as of those identified around strong \hi\ absorbers (LLSs and DLAs, $\log(N_{\rm HI}/{\rm cm^{-2}})\gtrsim17$) from \citet{lofthouse_muse_2023}. 

\begin{table*}[h]
\centering
\caption{Best-fit Schecther function parameters of the luminosity functions of LAEs associated with SBLAs identified in spectra with different $\Delta v_{\rm spec}$.}
\label{tab:LF_results}
\begin{tabular}{lccc}
\toprule
$\Delta v_{\rm spec}/(\rm km\,s^{-1})$ & $\alpha$ & $\log[L^\star/(\rm erg\,s^{-1})]$ & $\log[ \phi^\star/(\rm Mpc^{-3}\,dex^{-1})]$ \\
\midrule
$69$  & $-1.45 \pm 0.29$ & $42.60 \pm 0.19$ & $-2.83 \pm 0.30$ \\
$138$ & $-1.42 \pm 0.31$ & $42.60 \pm 0.20$ & $-2.52 \pm 0.31$ \\
$276$ & $-1.82 \pm 0.30$ & $42.87 \pm 0.37$ & $-2.72 \pm 0.61$ \\
\bottomrule
\end{tabular}
\tablefoot{Medians of the one-dimensional marginalized posterior distribution and the uncertainties represent the $16$th and $84$th percentiles.}
\end{table*}

We find that the luminosity functions of LAEs around SBLAs have a similar shape to those derived in MAGG, indicating no notable excess of either bright or faint LAEs. The main differences appear in the normalization, which traces the amplitude of the galaxy overdensity. This normalization increases with the spectral pixel size used to define the SBLAs, suggesting that $\Delta v_{\rm spec}$ plays a significant role in interpreting the nature of these absorption systems. In particular, SBLAs identified with $\Delta v_{\rm spec}=69\rm\,km\,s^{-1}$ show the lowest normalization, even lower than that of field LAEs. A possible scenario is that a significant fraction of these absorbers is pinpointing volumes that are not occupied by the LAEs detected in MAGG, a characteristic that could suggest a larger contribution from the IGM rather than the CGM at these velocity scales. This may also result due to contaminations in the SBLAs sample arising from noise fluctuations that can affect the purity of the selection for relatively high transmissions, $-0.05<F<0.25$. Selecting stronger absorbers (e.g. $-0.05<F<0.15$ and $-0.05<F<0.05$) mitigates this effect and results in normalizations that are up to $\approx0.25~{\rm dex}$ higher, strongly suggesting that the contribution of the CGM gas can be significant at these fluxes even for small $\Delta v_{\rm spec}$. Finally, a combination of this effect with the contribution of gas located on scales larger than the typical CGM can also be considered a plausible interpretation. For $\Delta v_{\rm spec}=138\rm\,km\,s^{-1}$, the luminosity function closely matches that of field LAEs and is thus consistent with SBLAs tracing the CGM of the galaxies. Finally, the strongest overdensity is observed for $\Delta v_{\rm spec}=276\rm\,km\,s^{-1}$, where the normalization exceeds the field, but is still significantly lower compared to the largest overdensity of LAEs observed around stronger \hi\ absorbers. These SBLAs may therefore trace regions populated by a larger number density of galaxies where it is more likely that a significant contribution to the observed absorption arises from overlapping halos. In such environments, the circumgalactic gas associated with neighboring systems can produce absorption components that are close in velocity space. If not individually resolved, their superimposition may appear as a single broader absorption feature.

\section{Discussion and conclusions}
\label{sec:disc}

The analysis of the link between LAEs and spectral bins at varying transmission in the MAGG fields shows that wavelength bins with $F<0.25$, selected as SBLAs, are more often found near LAEs than those at $F>0.25$. The correlation between SBLAs and galaxies also depends on the size of the spectral bin, being the strongest for $\Delta v_{\rm spec}=276\rm\,km\,s^{-1}$ and for small impact parameters, $R\lesssim 150\,\rm kpc$. From these results, we conclude that SBLAs identify regions of high optical depth that differ from random IGM regions, are preferentially occupied by star-forming galaxies, and do not overlap significantly with LLSs.

The findings of our LAE analysis share similarities with those of studies conducted in the vicinity of LBGs. \citet{Pieri2014} studied the associations with LBGs, finding that SBLAs with $F<0.25$ reside near an LBG $\approx 60~$percent of the time, in excess compared to a random distribution. Associations between LBGs and regions of enhanced absorption are also found in large spectroscopic surveys of galaxies in quasar fields. \citet{rudie_gaseous_2012} investigated the correlation between \ion{H}{I} absorbers in the range $10^{12}\le N_{\rm\hi}/{\rm cm^{-2}}\le10^{21}$ and LBGs with $z\approx2-3$.  Hydrogen absorbers with N$_{\rm\hi}/{\rm cm^{-2}}\gtrsim 10^{14.5}$ cluster preferentially within $\approx300~\rm kpc$ and $\approx300~\rm km~s^{-1}$ of galaxies, a trend that is not observed at a lower column density. These authors further estimate that circumgalactic gas can account for $\approx 50$~percent of all systems with N$_{\rm\hi}/{\rm cm^{-2}}\gtrsim10^{15.5}$, a value in line with the inferred column density of SBLAs \citep{Morrison2024}. Similar conclusions are reached by \citet{rakic_neutral_2012}, who examined the correlation between LBGs and \ion{H}{I} optical depth in the same sample of LBGs studied by \citet{rudie_gaseous_2012}. Regions with elevated optical depth, $\tau_{\rm Ly\alpha} > 0.1$, are found closer to galaxies than random regions. Circumgalactic gas within $\approx 100-200~\rm kpc$ contributes to absorption with optical depth $\tau_{\rm Ly\alpha}\gtrsim1$, corresponding to column densities N$_{\rm\hi}/{\rm cm^{-2}}\gtrsim 10^{14}$ and a transmitted flux similar to the one that defines SBLAs. 

Albeit not strictly quantitative, this comparison underscores similarities between LBGs and our findings with LAEs: $\approx 40-60$~percent of the SBLAs are identifiable with circumgalactic gas within $\approx 300~\rm kpc$ and $\approx 300~\rm km~s^{-1}$ of star-forming galaxies. Exploring stacked quasar spectra around $z\approx3-4$ LAEs, \citet{muzahid_musequbes_2021} found that the \hi\ optical depth is significantly higher within LOS separations of $\abs{\Delta\varv}\lesssim500\,{\rm km\,s^{-1}}$ from the galaxies compared to random regions in the universe \citep[see also][]{Matthee+2024}. In addition, they reported a peak at $F<0.05$ in the flux distribution of pixels within $\pm100\,{\rm km\,s^{-1}}$ from the LAEs, in agreement with the excess of low-transmission absorbers we observe in Fig.~\ref{fig:correlation_100kpc}.

LAEs and LBGs are, however, only partially overlapping populations: the typical halo mass of LBGs is inferred around $\approx 10^{11.5}-10^{12}~$M$_\odot$ \citep{bielby_vlt_2013,rakic_measurement_2013} and the mass of LAEs selected by typical IFS observations spans a luminosity-dependent range of $M_{\rm h}\approx 10^{10}-10^{11.5}{\rm M_\odot}$ \citep{herrero_alonso_clustering_2023}. Recent theoretical work by \citet{santos_finding_2025} suggests a connection between the velocity scale over which SBLAs are found and the halo mass of the associated galaxies. According to the cross-correlations in Fig.~\ref{fig:correlation_links}, stronger signal over velocity scales $\gtrsim 250~\rm km~s^{-1}$ is to be expected for halos with $M_{\rm h}\gtrsim 10^{11}~{\rm M_\odot}$, with a peak at $M_{\rm h}\approx 10^{11.5}~{\rm M_\odot}$. The LAEs in MAGG are bright enough ($\log[L_{\rm Ly\alpha}/({\rm erg~s^{-1}})]\gtrsim 41.8$; see Fig.~\ref{fig:LF}) to enter the intermediate ($\langle \log[L_{Ly\alpha}/({\rm erg\,s^{-1}})]\rangle=41.64$) and high-luminosity range ($\langle \log[L_{Ly\alpha}/({\rm erg\,s^{-1}})]\rangle=42.34$) according to the definition in the clustering analysis by \citet{herrero_alonso_clustering_2023}. The inferred mass is therefore around $M_{\rm h}\approx 10^{11.2}~{\rm M_\odot}$, which is just a factor $\approx 2$ smaller than the peak value predicted by \citet{santos_finding_2025}. Given the wide distributions reported in their figure~B.1, this small discrepancy does not imply a tension. Expanded samples should next explore the dependence of the LAE-SBLA association on luminosity, as a test of the predicted halo mass range. We also note that observational selection effects limit the detection of LAEs to relatively luminous galaxies. As a result, our analysis likely traces only a subset of the full galaxy population associated with SBLAs, namely the brighter \lya -emitting galaxies (with a 50\% completeness limit corresponding to $\log[L_{\rm Ly\alpha}/({\rm erg\,s^{-1}})]\approx41.87$, see \citealp{fossati_muse_2021, galbiati_muse_2023}) accessible to our survey. Other galaxies, including systems with weak or absent \lya\ emission, may remain undetected. Therefore, the lack of a detected association for the remaining absorbers does not necessarily imply the absence of circumgalactic gas.

In conclusion, our analysis confirms that a significant fraction ($\approx 40$~percent) of SBLAs is associated with circumgalactic gas in LAEs, reaffirming that strong blended absorbers can pinpoint the CGM of high redshift galaxies at intermediate column densities in between the Ly$\alpha$ forest and the optically-thick LLSs \citep{rudie_gaseous_2012, Morrison2024}. We also find that the LAE-SBLA cross-correlation is strongly sensitive to the spectral binning used to identify these absorbers, consistent with trends suggested by recent observational \citep{Perez2023, Morrison2024} and theoretical \citep{santos_finding_2025} studies. Thus, the current framework appears to be effective in linking absorbers to halos and galaxies, although it has only been tested at $z\lesssim3$. On the other hand, further work is needed to quantify how the choice of the selection parameters influences the SBLA-halo association and, in particular, there remains room for testing how optimizing the flux threshold as a function of both redshift and spectral binning would improve the selection and characterization of SBLAs, especially at $z\gtrsim3$.


\begin{acknowledgements} 
We thank the anonymous referee for carefully reading the paper and for providing useful suggestions that helped improve the manuscript. This paper uses data from the MAGG survey, based on observations collected at the European Organisation for Astronomical Research in the Southern Hemisphere under ESO programme IDs 197.A-0384, 065.O-0299, 067.A-0022, 068.A-0461, 068.A-0492, 068.A-0600, 068.B-0115, 069.A-0613, 071.A-0067, 071.A-0114, 073.A-0071, 073.A-0653, 073.B-0787, 074.A-0306, 075.A-0464, 077.A-0166, 080.A-0482, 083.A-0042, 091.A-0833, 092.A-0011, 093.A-0575, 094.A-0280, 094.A-0131, 094.A-0585, 095.A-0200, 096.A-0937, 097.A-0089, 099.A-0159, 166.A-0106, 189.A-0424, 0109.A-0559. This study is supported by the Italian Ministry for Research and University (MUR) under Grant 'Progetto Dipartimenti di Eccellenza 2023-2027' (BiCoQ). MG also acknowledges support by Progetto FARE 2020 {\it Svelare i nodi massicci della CosmicWeb} ID 2021-NAZ-0326/PER. This research used Astropy\footnote{\url{http://www.astropy.org}}, a community-developed core Python package for Astronomy \citep{astropy:2013,astropy:2018,astropy:2022}, NumPy \citep{harris_array_2020}, SciPy \citep{virtanen_scipy_2020}, Matplotlib \citep{hunter_matplotlib_2007}.
\end{acknowledgements}


\bibliographystyle{aa}
\bibliography{reference}

@article{hunter_matplotlib_2007,
	title = {Matplotlib: {A} {2D} {Graphics} {Environment}},
	volume = {9},
	copyright = {https://ieeexplore.ieee.org/Xplorehelp/downloads/license-information/IEEE.html},
	issn = {1521-9615},
	shorttitle = {Matplotlib},
	url = {http://ieeexplore.ieee.org/document/4160265/},
	doi = {10.1109/MCSE.2007.55},
	number = {3},
	urldate = {2024-08-03},
	journal = {Computing in Science \& Engineering},
	author = {Hunter, John D.},
	year = {2007},
	pages = {90--95},
}

@article{virtanen_scipy_2020,
	title = {{SciPy} 1.0: fundamental algorithms for scientific computing in {Python}},
	volume = {17},
	issn = {1548-7091, 1548-7105},
	shorttitle = {{SciPy} 1.0},
	url = {https://www.nature.com/articles/s41592-019-0686-2},
	doi = {10.1038/s41592-019-0686-2},
	language = {en},
	number = {3},
	urldate = {2024-08-03},
	journal = {Nature Methods},
	author = {Virtanen, Pauli and Gommers, Ralf and Oliphant, Travis E. and Haberland, Matt and Reddy, Tyler and Cournapeau, David and Burovski, Evgeni and Peterson, Pearu and Weckesser, Warren and Bright, Jonathan and Van Der Walt, Stéfan J. and Brett, Matthew and Wilson, Joshua and Millman, K. Jarrod and Mayorov, Nikolay and Nelson, Andrew R. J. and Jones, Eric and Kern, Robert and Larson, Eric and Carey, C J and Polat, İlhan and Feng, Yu and Moore, Eric W. and VanderPlas, Jake and Laxalde, Denis and Perktold, Josef and Cimrman, Robert and Henriksen, Ian and Quintero, E. A. and Harris, Charles R. and Archibald, Anne M. and Ribeiro, Antônio H. and Pedregosa, Fabian and Van Mulbregt, Paul and {SciPy 1.0 Contributors} and Vijaykumar, Aditya and Bardelli, Alessandro Pietro and Rothberg, Alex and Hilboll, Andreas and Kloeckner, Andreas and Scopatz, Anthony and Lee, Antony and Rokem, Ariel and Woods, C. Nathan and Fulton, Chad and Masson, Charles and Häggström, Christian and Fitzgerald, Clark and Nicholson, David A. and Hagen, David R. and Pasechnik, Dmitrii V. and Olivetti, Emanuele and Martin, Eric and Wieser, Eric and Silva, Fabrice and Lenders, Felix and Wilhelm, Florian and Young, G. and Price, Gavin A. and Ingold, Gert-Ludwig and Allen, Gregory E. and Lee, Gregory R. and Audren, Hervé and Probst, Irvin and Dietrich, Jörg P. and Silterra, Jacob and Webber, James T and Slavič, Janko and Nothman, Joel and Buchner, Johannes and Kulick, Johannes and Schönberger, Johannes L. and De Miranda Cardoso, José Vinícius and Reimer, Joscha and Harrington, Joseph and Rodríguez, Juan Luis Cano and Nunez-Iglesias, Juan and Kuczynski, Justin and Tritz, Kevin and Thoma, Martin and Newville, Matthew and Kümmerer, Matthias and Bolingbroke, Maximilian and Tartre, Michael and Pak, Mikhail and Smith, Nathaniel J. and Nowaczyk, Nikolai and Shebanov, Nikolay and Pavlyk, Oleksandr and Brodtkorb, Per A. and Lee, Perry and McGibbon, Robert T. and Feldbauer, Roman and Lewis, Sam and Tygier, Sam and Sievert, Scott and Vigna, Sebastiano and Peterson, Stefan and More, Surhud and Pudlik, Tadeusz and Oshima, Takuya and Pingel, Thomas J. and Robitaille, Thomas P. and Spura, Thomas and Jones, Thouis R. and Cera, Tim and Leslie, Tim and Zito, Tiziano and Krauss, Tom and Upadhyay, Utkarsh and Halchenko, Yaroslav O. and Vázquez-Baeza, Yoshiki},
	month = mar,
	year = {2020},
	pages = {261--272},
}

@article{harris_array_2020,
	title = {Array programming with {NumPy}},
	volume = {585},
	issn = {0028-0836, 1476-4687},
	url = {https://www.nature.com/articles/s41586-020-2649-2},
	doi = {10.1038/s41586-020-2649-2},
	language = {en},
	number = {7825},
	urldate = {2024-08-03},
	journal = {Nature},
	author = {Harris, Charles R. and Millman, K. Jarrod and Van Der Walt, Stéfan J. and Gommers, Ralf and Virtanen, Pauli and Cournapeau, David and Wieser, Eric and Taylor, Julian and Berg, Sebastian and Smith, Nathaniel J. and Kern, Robert and Picus, Matti and Hoyer, Stephan and Van Kerkwijk, Marten H. and Brett, Matthew and Haldane, Allan and Del Río, Jaime Fernández and Wiebe, Mark and Peterson, Pearu and Gérard-Marchant, Pierre and Sheppard, Kevin and Reddy, Tyler and Weckesser, Warren and Abbasi, Hameer and Gohlke, Christoph and Oliphant, Travis E.},
	month = sep,
	year = {2020},
	pages = {357--362},
}

@article{astropy:2013,
Adsurl = {http://adsabs.harvard.edu/abs/2013A%26A...558A..33A},
Archiveprefix = {arXiv},
Author = {{Astropy Collaboration} and {Robitaille}, T.~P. and {Tollerud}, E.~J. and {Greenfield}, P. and {Droettboom}, M. and {Bray}, E. and {Aldcroft}, T. and {Davis}, M. and {Ginsburg}, A. and {Price-Whelan}, A.~M. and {Kerzendorf}, W.~E. and {Conley}, A. and {Crighton}, N. and {Barbary}, K. and {Muna}, D. and {Ferguson}, H. and {Grollier}, F. and {Parikh}, M.~M. and {Nair}, P.~H. and {Unther}, H.~M. and {Deil}, C. and {Woillez}, J. and {Conseil}, S. and {Kramer}, R. and {Turner}, J.~E.~H. and {Singer}, L. and {Fox}, R. and {Weaver}, B.~A. and {Zabalza}, V. and {Edwards}, Z.~I. and {Azalee Bostroem}, K. and {Burke}, D.~J. and {Casey}, A.~R. and {Crawford}, S.~M. and {Dencheva}, N. and {Ely}, J. and {Jenness}, T. and {Labrie}, K. and {Lim}, P.~L. and {Pierfederici}, F. and {Pontzen}, A. and {Ptak}, A. and {Refsdal}, B. and {Servillat}, M. and {Streicher}, O.},
Doi = {10.1051/0004-6361/201322068},
Eid = {A33},
Eprint = {1307.6212},
Journal = {\aap},
Month = oct,
Pages = {A33},
Primaryclass = {astro-ph.IM},
Title = {{Astropy: A community Python package for astronomy}},
Volume = 558,
Year = 2013}

@ARTICLE{astropy:2018,
       author = {{Astropy Collaboration} and {Price-Whelan}, A.~M. and
         {Sip{\H{o}}cz}, B.~M. and {G{\"u}nther}, H.~M. and {Lim}, P.~L. and
         {Crawford}, S.~M. and {Conseil}, S. and {Shupe}, D.~L. and
         {Craig}, M.~W. and {Dencheva}, N. and {Ginsburg}, A. and {Vand
        erPlas}, J.~T. and {Bradley}, L.~D. and {P{\'e}rez-Su{\'a}rez}, D. and
         {de Val-Borro}, M. and {Aldcroft}, T.~L. and {Cruz}, K.~L. and
         {Robitaille}, T.~P. and {Tollerud}, E.~J. and {Ardelean}, C. and
         {Babej}, T. and {Bach}, Y.~P. and {Bachetti}, M. and {Bakanov}, A.~V. and
         {Bamford}, S.~P. and {Barentsen}, G. and {Barmby}, P. and
         {Baumbach}, A. and {Berry}, K.~L. and {Biscani}, F. and {Boquien}, M. and
         {Bostroem}, K.~A. and {Bouma}, L.~G. and {Brammer}, G.~B. and
         {Bray}, E.~M. and {Breytenbach}, H. and {Buddelmeijer}, H. and
         {Burke}, D.~J. and {Calderone}, G. and {Cano Rodr{\'\i}guez}, J.~L. and
         {Cara}, M. and {Cardoso}, J.~V.~M. and {Cheedella}, S. and {Copin}, Y. and
         {Corrales}, L. and {Crichton}, D. and {D'Avella}, D. and {Deil}, C. and
         {Depagne}, {\'E}. and {Dietrich}, J.~P. and {Donath}, A. and
         {Droettboom}, M. and {Earl}, N. and {Erben}, T. and {Fabbro}, S. and
         {Ferreira}, L.~A. and {Finethy}, T. and {Fox}, R.~T. and
         {Garrison}, L.~H. and {Gibbons}, S.~L.~J. and {Goldstein}, D.~A. and
         {Gommers}, R. and {Greco}, J.~P. and {Greenfield}, P. and
         {Groener}, A.~M. and {Grollier}, F. and {Hagen}, A. and {Hirst}, P. and
         {Homeier}, D. and {Horton}, A.~J. and {Hosseinzadeh}, G. and {Hu}, L. and
         {Hunkeler}, J.~S. and {Ivezi{\'c}}, {\v{Z}}. and {Jain}, A. and
         {Jenness}, T. and {Kanarek}, G. and {Kendrew}, S. and {Kern}, N.~S. and
         {Kerzendorf}, W.~E. and {Khvalko}, A. and {King}, J. and {Kirkby}, D. and
         {Kulkarni}, A.~M. and {Kumar}, A. and {Lee}, A. and {Lenz}, D. and
         {Littlefair}, S.~P. and {Ma}, Z. and {Macleod}, D.~M. and
         {Mastropietro}, M. and {McCully}, C. and {Montagnac}, S. and
         {Morris}, B.~M. and {Mueller}, M. and {Mumford}, S.~J. and {Muna}, D. and
         {Murphy}, N.~A. and {Nelson}, S. and {Nguyen}, G.~H. and
         {Ninan}, J.~P. and {N{\"o}the}, M. and {Ogaz}, S. and {Oh}, S. and
         {Parejko}, J.~K. and {Parley}, N. and {Pascual}, S. and {Patil}, R. and
         {Patil}, A.~A. and {Plunkett}, A.~L. and {Prochaska}, J.~X. and
         {Rastogi}, T. and {Reddy Janga}, V. and {Sabater}, J. and
         {Sakurikar}, P. and {Seifert}, M. and {Sherbert}, L.~E. and
         {Sherwood-Taylor}, H. and {Shih}, A.~Y. and {Sick}, J. and
         {Silbiger}, M.~T. and {Singanamalla}, S. and {Singer}, L.~P. and
         {Sladen}, P.~H. and {Sooley}, K.~A. and {Sornarajah}, S. and
         {Streicher}, O. and {Teuben}, P. and {Thomas}, S.~W. and
         {Tremblay}, G.~R. and {Turner}, J.~E.~H. and {Terr{\'o}n}, V. and
         {van Kerkwijk}, M.~H. and {de la Vega}, A. and {Watkins}, L.~L. and
         {Weaver}, B.~A. and {Whitmore}, J.~B. and {Woillez}, J. and
         {Zabalza}, V. and {Astropy Contributors}},
        title = "{The Astropy Project: Building an Open-science Project and Status of the v2.0 Core Package}",
      journal = {\aj},
         year = 2018,
        month = sep,
       volume = {156},
       number = {3},
          eid = {123},
        pages = {123},
          doi = {10.3847/1538-3881/aabc4f},
archivePrefix = {arXiv},
       eprint = {1801.02634},
 primaryClass = {astro-ph.IM},
       adsurl = {https://ui.adsabs.harvard.edu/abs/2018AJ....156..123A}
}

@ARTICLE{astropy:2022,
       author = {{Astropy Collaboration} and {Price-Whelan}, Adrian M. and {Lim}, Pey Lian and {Earl}, Nicholas and {Starkman}, Nathaniel and {Bradley}, Larry and {Shupe}, David L. and {Patil}, Aarya A. and {Corrales}, Lia and {Brasseur}, C.~E. and {N{"o}the}, Maximilian and {Donath}, Axel and {Tollerud}, Erik and {Morris}, Brett M. and {Ginsburg}, Adam and {Vaher}, Eero and {Weaver}, Benjamin A. and {Tocknell}, James and {Jamieson}, William and {van Kerkwijk}, Marten H. and {Robitaille}, Thomas P. and {Merry}, Bruce and {Bachetti}, Matteo and {G{"u}nther}, H. Moritz and {Aldcroft}, Thomas L. and {Alvarado-Montes}, Jaime A. and {Archibald}, Anne M. and {B{'o}di}, Attila and {Bapat}, Shreyas and {Barentsen}, Geert and {Baz{'a}n}, Juanjo and {Biswas}, Manish and {Boquien}, M{'e}d{'e}ric and {Burke}, D.~J. and {Cara}, Daria and {Cara}, Mihai and {Conroy}, Kyle E. and {Conseil}, Simon and {Craig}, Matthew W. and {Cross}, Robert M. and {Cruz}, Kelle L. and {D'Eugenio}, Francesco and {Dencheva}, Nadia and {Devillepoix}, Hadrien A.~R. and {Dietrich}, J{"o}rg P. and {Eigenbrot}, Arthur Davis and {Erben}, Thomas and {Ferreira}, Leonardo and {Foreman-Mackey}, Daniel and {Fox}, Ryan and {Freij}, Nabil and {Garg}, Suyog and {Geda}, Robel and {Glattly}, Lauren and {Gondhalekar}, Yash and {Gordon}, Karl D. and {Grant}, David and {Greenfield}, Perry and {Groener}, Austen M. and {Guest}, Steve and {Gurovich}, Sebastian and {Handberg}, Rasmus and {Hart}, Akeem and {Hatfield-Dodds}, Zac and {Homeier}, Derek and {Hosseinzadeh}, Griffin and {Jenness}, Tim and {Jones}, Craig K. and {Joseph}, Prajwel and {Kalmbach}, J. Bryce and {Karamehmetoglu}, Emir and {Ka{l}uszy{'n}ski}, Miko{l}aj and {Kelley}, Michael S.~P. and {Kern}, Nicholas and {Kerzendorf}, Wolfgang E. and {Koch}, Eric W. and {Kulumani}, Shankar and {Lee}, Antony and {Ly}, Chun and {Ma}, Zhiyuan and {MacBride}, Conor and {Maljaars}, Jakob M. and {Muna}, Demitri and {Murphy}, N.~A. and {Norman}, Henrik and {O'Steen}, Richard and {Oman}, Kyle A. and {Pacifici}, Camilla and {Pascual}, Sergio and {Pascual-Granado}, J. and {Patil}, Rohit R. and {Perren}, Gabriel I. and {Pickering}, Timothy E. and {Rastogi}, Tanuj and {Roulston}, Benjamin R. and {Ryan}, Daniel F. and {Rykoff}, Eli S. and {Sabater}, Jose and {Sakurikar}, Parikshit and {Salgado}, Jes{'u}s and {Sanghi}, Aniket and {Saunders}, Nicholas and {Savchenko}, Volodymyr and {Schwardt}, Ludwig and {Seifert-Eckert}, Michael and {Shih}, Albert Y. and {Jain}, Anany Shrey and {Shukla}, Gyanendra and {Sick}, Jonathan and {Simpson}, Chris and {Singanamalla}, Sudheesh and {Singer}, Leo P. and {Singhal}, Jaladh and {Sinha}, Manodeep and {Sip{H{o}}cz}, Brigitta M. and {Spitler}, Lee R. and {Stansby}, David and {Streicher}, Ole and {{{S}}umak}, Jani and {Swinbank}, John D. and {Taranu}, Dan S. and {Tewary}, Nikita and {Tremblay}, Grant R. and {Val-Borro}, Miguel de and {Van Kooten}, Samuel J. and {Vasovi{'c}}, Zlatan and {Verma}, Shresth and {de Miranda Cardoso}, Jos{'e} Vin{'i}cius and {Williams}, Peter K.~G. and {Wilson}, Tom J. and {Winkel}, Benjamin and {Wood-Vasey}, W.~M. and {Xue}, Rui and {Yoachim}, Peter and {Zhang}, Chen and {Zonca}, Andrea and {Astropy Project Contributors}},
        title = "{The Astropy Project: Sustaining and Growing a Community-oriented Open-source Project and the Latest Major Release (v5.0) of the Core Package}",
      journal = {\apj},
         year = 2022,
        month = aug,
       volume = {935},
       number = {2},
          eid = {167},
        pages = {167},
          doi = {10.3847/1538-4357/ac7c74},
archivePrefix = {arXiv},
       eprint = {2206.14220},
 primaryClass = {astro-ph.IM},
       adsurl = {https://ui.adsabs.harvard.edu/abs/2022ApJ...935..167A}
}

@article{dutta_metal-enriched_2021,
	title = {Metal-enriched halo gas across galaxy overdensities over the last 10 billion years},
	volume = {508},
	issn = {0035-8711},
	url = {https://ui.adsabs.harvard.edu/abs/2021MNRAS.508.4573D},
	doi = {10.1093/mnras/stab2752},
	journal = {Monthly Notices of the Royal Astronomical Society},
	author = {Dutta, Rajeshwari and Fumagalli, Michele and Fossati, Matteo and Bielby, Richard M. and Stott, John P. and Lofthouse, Emma K. and Cantalupo, Sebastiano and Cullen, Fergus and Crain, Robert A. and Tripp, Todd M. and Prochaska, J. Xavier and Arrigoni Battaia, Fabrizio and Burchett, Joseph N. and Fynbo, Johan P. U. and Murphy, Michael T. and Schaye, Joop and Tejos, Nicolas and Theuns, Tom},
	month = dec,
	year = {2021},
	note = {Publisher: OUP},
	pages = {4573--4599},
}

@ARTICLE{Ouchi2020,
   author = {{Ouchi}, M. and {Ono}, Y. and {Shibuya}, T.},
    title = "{The Physics of Galaxy Formation: A Global, Statistical View on the Cooling and Contraction Phases of the High-Redshift Galaxies}",
  journal = {\araa},
     year = 2020,
    month = sep,
   volume = 58,
    pages = {617-652},
      doi = {10.1146/annurev-astro-032620-021859},
   adsurl = {https://ui.adsabs.harvard.edu/abs/2020ARA&A..58..617O}
}

@ARTICLE{Schlafly2011,
   author = {{Schlafly}, E.~F. and {Finkbeiner}, D.~P.},
    title = "{Measuring Reddening with Sloan Digital Sky Survey Stellar Spectra and Recalibrating SFD}",
  journal = {\apj},
archivePrefix = "arXiv",
   eprint = {1012.4804},
     year = 2011,
    month = aug,
   volume = 737,
    pages = {103},
      doi = {10.1088/0004-637X/737/2/103},
   adsurl = {https://ui.adsabs.harvard.edu/abs/2011ApJ...737..103S}
}

@ARTICLE{Cantalupo2019,
       author = {{Cantalupo}, Sebastiano and {Pezzulli}, Gabriele and {Lilly}, Simon J. and {Marino}, Raffaella Anna and {Gallego}, Sofia G. and {Schaye}, Joop and {Bacon}, Roland and {Feltre}, Anna and {Kollatschny}, Wolfram and {Nanayakkara}, Themiya and {Richard}, Johan and {Wendt}, Martin and {Wisotzki}, Lutz and {Prochaska}, J. Xavier},
        title = "{The large- and small-scale properties of the intergalactic gas in the Slug Ly {\ensuremath{\alpha}} nebula revealed by MUSE He II emission observations}",
      journal = {\mnras},
         year = 2019,
        month = mar,
       volume = {483},
       number = {4},
        pages = {5188-5204},
          doi = {10.1093/mnras/sty3481},
archivePrefix = {arXiv},
       eprint = {1811.11783},
 primaryClass = {astro-ph.GA},
       adsurl = {https://ui.adsabs.harvard.edu/abs/2019MNRAS.483.5188C}
}

@ARTICLE{Hinton2016,
       author = {{Hinton}, S.~R. and {Davis}, Tamara M. and {Lidman}, C. and {Glazebrook}, K. and {Lewis}, G.~F.},
        title = "{MARZ: Manual and automatic redshifting software}",
      journal = {Astronomy and Computing},
         year = 2016,
        month = apr,
       volume = {15},
        pages = {61-71},
          doi = {10.1016/j.ascom.2016.03.001},
archivePrefix = {arXiv},
       eprint = {1603.09438},
 primaryClass = {astro-ph.IM},
       adsurl = {https://ui.adsabs.harvard.edu/abs/2016A&C....15...61H}
}

@ARTICLE{Carnall2017,
       author = {{Carnall}, A.~C.},
        title = "{SpectRes: A Fast Spectral Resampling Tool in Python}",
      journal = {arXiv e-prints},
         year = 2017,
        month = may,
          eid = {arXiv:1705.05165},
        pages = {arXiv:1705.05165},
          doi = {10.48550/arXiv.1705.05165},
archivePrefix = {arXiv},
       eprint = {1705.05165},
 primaryClass = {astro-ph.IM},
       adsurl = {https://ui.adsabs.harvard.edu/abs/2017arXiv170505165C}
}

@ARTICLE{Bertin1996,
   author = {{Bertin}, E. and {Arnouts}, S.},
    title = "{SExtractor: Software for source extraction}",
  journal = {\aaps},
     year = 1996,
    month = nov,
   volume = 117,
    pages = {393-404},
      doi = {10.1051/aas:1996164},
   adsurl = {https://ui.adsabs.harvard.edu/abs/1996A%26AS..117..393B}
}

@ARTICLE{Fitzpatrick1999,
   author = {{Fitzpatrick}, E.~L.},
    title = "{Correcting for the Effects of Interstellar Extinction}",
  journal = {\pasp},
     year = 1999,
    month = jan,
   volume = 111,
    pages = {63-75},
      doi = {10.1086/316293},
   adsurl = {https://ui.adsabs.harvard.edu/abs/1999PASP..111...63F}
}

@ARTICLE{galbiati_muse_2024,
       author = {{Galbiati}, Marta and {Dutta}, Rajeshwari and {Fumagalli}, Michele and {Fossati}, Matteo and {Cantalupo}, Sebastiano},
        title = "{MUSE Analysis of Gas around Galaxies (MAGG): VI. The cool and enriched gas environment of z {\ensuremath{\gtrsim}} 3 Ly{\ensuremath{\alpha}} emitters}",
      journal = {\aap},
         year = 2024,
        month = oct,
       volume = {690},
          eid = {A7},
        pages = {A7},
          doi = {10.1051/0004-6361/202450741},
archivePrefix = {arXiv},
       eprint = {2406.10350},
 primaryClass = {astro-ph.GA},
       adsurl = {https://ui.adsabs.harvard.edu/abs/2024A&A...690A...7G}
}

@ARTICLE{galbiati_muse_2023,
       author = {{Galbiati}, Marta and {Fumagalli}, Michele and {Fossati}, Matteo and {Lofthouse}, Emma K. and {Dutta}, Rajeshwari and {Prochaska}, J. Xavier and {Murphy}, Michael T. and {Cantalupo}, Sebastiano},
        title = "{MUSE Analysis of Gas around Galaxies (MAGG) - V. Linking ionized gas traced by C IV and Si IV absorbers to Ly {\ensuremath{\alpha}} emitting galaxies at z {\ensuremath{\approx}} 3.0-4.5}",
      journal = {\mnras},
         year = 2023,
        month = sep,
       volume = {524},
       number = {3},
        pages = {3474-3501},
          doi = {10.1093/mnras/stad2087},
archivePrefix = {arXiv},
       eprint = {2302.00021},
 primaryClass = {astro-ph.GA},
       adsurl = {https://ui.adsabs.harvard.edu/abs/2023MNRAS.524.3474G}
}

@article{lofthouse_muse_2023,
       author = {{Lofthouse}, Emma K. and {Fumagalli}, Michele and {Fossati}, Matteo and {Dutta}, Rajeshwari and {Galbiati}, Marta and {Arrigoni Battaia}, Fabrizio and {Cantalupo}, Sebastiano and {Christensen}, Lise and {Cooke}, Ryan J. and {Longobardi}, Alessia and {Murphy}, Michael T. and {Prochaska}, J. Xavier},
        title = "{MUSE Analysis of Gas around Galaxies (MAGG) - IV. The gaseous environment of z   3-4 Ly {\ensuremath{\alpha}} emitting galaxies}",
      journal = {\mnras},
         year = 2023,
        month = jan,
       volume = {518},
       number = {1},
        pages = {305-331},
          doi = {10.1093/mnras/stac3089},
archivePrefix = {arXiv},
       eprint = {2209.15021},
 primaryClass = {astro-ph.GA},
       adsurl = {https://ui.adsabs.harvard.edu/abs/2023MNRAS.518..305L}
}

@article{fossati_muse_2021,
       author = {{Fossati}, M. and {Fumagalli}, M. and {Lofthouse}, E.~K. and {Dutta}, R. and {Cantalupo}, S. and {Arrigoni Battaia}, F. and {Fynbo}, J.~P.~U. and {Lusso}, E. and {Murphy}, M.~T. and {Prochaska}, J.~X. and {Theuns}, T. and {Cooke}, R.~J.},
        title = "{MUSE analysis of gas around galaxies (MAGG) - III. The gas and galaxy environment of z = 3-4.5 quasars}",
      journal = {\mnras},
         year = 2021,
        month = may,
       volume = {503},
       number = {2},
        pages = {3044-3064},
          doi = {10.1093/mnras/stab660},
archivePrefix = {arXiv},
       eprint = {2103.01960},
 primaryClass = {astro-ph.GA},
       adsurl = {https://ui.adsabs.harvard.edu/abs/2021MNRAS.503.3044F}
}

@article{dutta_muse_2020,
       author = {{Dutta}, Rajeshwari and {Fumagalli}, Michele and {Fossati}, Matteo and {Lofthouse}, Emma K. and {Prochaska}, J. Xavier and {Arrigoni Battaia}, Fabrizio and {Bielby}, Richard M. and {Cantalupo}, Sebastiano and {Cooke}, Ryan J. and {Murphy}, Michael T. and {O'Meara}, John M.},
        title = "{MUSE Analysis of Gas around Galaxies (MAGG) - II: metal-enriched halo gas around z {\ensuremath{\sim}} 1 galaxies}",
      journal = {\mnras},
         year = 2020,
        month = dec,
       volume = {499},
       number = {4},
        pages = {5022-5046},
          doi = {10.1093/mnras/staa3147},
archivePrefix = {arXiv},
       eprint = {2009.14219},
 primaryClass = {astro-ph.GA},
       adsurl = {https://ui.adsabs.harvard.edu/abs/2020MNRAS.499.5022D}
}

@article{lofthouse_muse_2020,
       author = {{Lofthouse}, Emma K. and {Fumagalli}, Michele and {Fossati}, Matteo and {O'Meara}, John M. and {Murphy}, Michael T. and {Christensen}, Lise and {Prochaska}, J. Xavier and {Cantalupo}, Sebastiano and {Bielby}, Richard M. and {Cooke}, Ryan J. and {Lusso}, Elisabeta and {Morris}, Simon L.},
        title = "{MUSE Analysis of Gas around Galaxies (MAGG) - I: Survey design and the environment of a near pristine gas cloud at z {\ensuremath{\approx}} 3.5}",
      journal = {\mnras},
         year = 2020,
        month = jan,
       volume = {491},
       number = {2},
        pages = {2057-2074},
          doi = {10.1093/mnras/stz3066},
archivePrefix = {arXiv},
       eprint = {1910.13458},
 primaryClass = {astro-ph.GA},
       adsurl = {https://ui.adsabs.harvard.edu/abs/2020MNRAS.491.2057L}
}

@INPROCEEDINGS{bacon_muse_2010,
       author = {{Bacon}, R. and {Accardo}, M. and {Adjali}, L. and {Anwand}, H. and {Bauer}, S. and {Biswas}, I. and {Blaizot}, J. and {Boudon}, D. and {Brau-Nogue}, S. and {Brinchmann}, J. and {Caillier}, P. and {Capoani}, L. and {Carollo}, C.~M. and {Contini}, T. and {Couderc}, P. and {Daguis{\'e}}, E. and {Deiries}, S. and {Delabre}, B. and {Dreizler}, S. and {Dubois}, J. and {Dupieux}, M. and {Dupuy}, C. and {Emsellem}, E. and {Fechner}, T. and {Fleischmann}, A. and {Fran{\c{c}}ois}, M. and {Gallou}, G. and {Gharsa}, T. and {Glindemann}, A. and {Gojak}, D. and {Guiderdoni}, B. and {Hansali}, G. and {Hahn}, T. and {Jarno}, A. and {Kelz}, A. and {Koehler}, C. and {Kosmalski}, J. and {Laurent}, F. and {Le Floch}, M. and {Lilly}, S.~J. and {Lizon}, J. -L. and {Loupias}, M. and {Manescau}, A. and {Monstein}, C. and {Nicklas}, H. and {Olaya}, J. -C. and {Pares}, L. and {Pasquini}, L. and {P{\'e}contal-Rousset}, A. and {Pell{\'o}}, R. and {Petit}, C. and {Popow}, E. and {Reiss}, R. and {Remillieux}, A. and {Renault}, E. and {Roth}, M. and {Rupprecht}, G. and {Serre}, D. and {Schaye}, J. and {Soucail}, G. and {Steinmetz}, M. and {Streicher}, O. and {Stuik}, R. and {Valentin}, H. and {Vernet}, J. and {Weilbacher}, P. and {Wisotzki}, L. and {Yerle}, N.},
        title = "{The MUSE second-generation VLT instrument}",
    booktitle = {Ground-based and Airborne Instrumentation for Astronomy III},
         year = 2010,
       editor = {{McLean}, Ian S. and {Ramsay}, Suzanne K. and {Takami}, Hideki},
       series = {Society of Photo-Optical Instrumentation Engineers (SPIE) Conference Series},
       volume = {7735},
        month = jul,
          eid = {773508},
        pages = {773508},
          doi = {10.1117/12.856027},
archivePrefix = {arXiv},
       eprint = {2211.16795},
 primaryClass = {astro-ph.IM},
       adsurl = {https://ui.adsabs.harvard.edu/abs/2010SPIE.7735E..08B}
}

@article{tumlinson_circumgalactic_2017,
       author = {{Tumlinson}, Jason and {Peeples}, Molly S. and {Werk}, Jessica K.},
        title = "{The Circumgalactic Medium}",
      journal = {\araa},
         year = 2017,
        month = aug,
       volume = {55},
       number = {1},
        pages = {389-432},
          doi = {10.1146/annurev-astro-091916-055240},
archivePrefix = {arXiv},
       eprint = {1709.09180},
 primaryClass = {astro-ph.GA},
       adsurl = {https://ui.adsabs.harvard.edu/abs/2017ARA&A..55..389T}
}

@article{prochaska_definitive_2010,
       author = {{Prochaska}, J. Xavier and {O'Meara}, John M. and {Worseck}, Gabor},
        title = "{A Definitive Survey for Lyman Limit Systems at z \raisebox{-0.5ex}\textasciitilde 3.5 with the Sloan Digital Sky Survey}",
      journal = {\apj},
         year = 2010,
        month = jul,
       volume = {718},
       number = {1},
        pages = {392-416},
          doi = {10.1088/0004-637X/718/1/392},
archivePrefix = {arXiv},
       eprint = {0912.0292},
 primaryClass = {astro-ph.CO},
       adsurl = {https://ui.adsabs.harvard.edu/abs/2010ApJ...718..392P}
}

@article{khare_nature_2007,
       author = {{Khare}, P. and {Kulkarni}, V.~P. and {P{\'e}roux}, C. and {York}, D.~G. and {Lauroesch}, J.~T. and {Meiring}, J.~D.},
        title = "{The nature of damped Lyman {\ensuremath{\alpha}} and sub-damped Lyman {\ensuremath{\alpha}} absorbers}",
      journal = {\aap},
         year = 2007,
        month = mar,
       volume = {464},
       number = {2},
        pages = {487-493},
          doi = {10.1051/0004-6361:20066186},
archivePrefix = {arXiv},
       eprint = {astro-ph/0608127},
 primaryClass = {astro-ph},
       adsurl = {https://ui.adsabs.harvard.edu/abs/2007A&A...464..487K}
}

@article{rahmati_evolution_2013,
       author = {{Rahmati}, Alireza and {Pawlik}, Andreas H. and {Rai{\v{c}}evi{\'c}}, Milan and {Schaye}, Joop},
        title = "{On the evolution of the H I column density distribution in cosmological simulations}",
      journal = {\mnras},
         year = 2013,
        month = apr,
       volume = {430},
       number = {3},
        pages = {2427-2445},
          doi = {10.1093/mnras/stt066},
archivePrefix = {arXiv},
       eprint = {1210.7808},
 primaryClass = {astro-ph.CO},
       adsurl = {https://ui.adsabs.harvard.edu/abs/2013MNRAS.430.2427R}
}

@article{rauch_lyman_1998,
       author = {{Rauch}, Michael},
        title = "{The Lyman Alpha Forest in the Spectra of QSOs}",
      journal = {\araa},
         year = 1998,
        month = jan,
       volume = {36},
        pages = {267-316},
          doi = {10.1146/annurev.astro.36.1.267},
archivePrefix = {arXiv},
       eprint = {astro-ph/9806286},
 primaryClass = {astro-ph},
       adsurl = {https://ui.adsabs.harvard.edu/abs/1998ARA&A..36..267R}
}

@article{fumagalli_physical_2016,
       author = {{Fumagalli}, Michele and {O'Meara}, John M. and {Prochaska}, J. Xavier},
        title = "{The physical properties of z > 2 Lyman limit systems: new constraints for feedback and accretion models}",
      journal = {\mnras},
         year = 2016,
        month = feb,
       volume = {455},
       number = {4},
        pages = {4100-4121},
          doi = {10.1093/mnras/stv2616},
archivePrefix = {arXiv},
       eprint = {1511.01898},
 primaryClass = {astro-ph.GA},
       adsurl = {https://ui.adsabs.harvard.edu/abs/2016MNRAS.455.4100F}
}

@article{fumagalli_dissecting_2013,
       author = {{Fumagalli}, Michele and {O'Meara}, John M. and {Prochaska}, J. Xavier and {Worseck}, Gabor},
        title = "{Dissecting the Properties of Optically Thick Hydrogen at the Peak of Cosmic Star Formation History}",
      journal = {\apj},
         year = 2013,
        month = sep,
       volume = {775},
       number = {1},
          eid = {78},
        pages = {78},
          doi = {10.1088/0004-637X/775/1/78},
archivePrefix = {arXiv},
       eprint = {1308.1101},
 primaryClass = {astro-ph.CO},
       adsurl = {https://ui.adsabs.harvard.edu/abs/2013ApJ...775...78F}
}

@article{fumagalli_absorption-line_2011,
       author = {{Fumagalli}, Michele and {Prochaska}, J. Xavier and {Kasen}, Daniel and {Dekel}, Avishai and {Ceverino}, Daniel and {Primack}, Joel R.},
        title = "{Absorption-line systems in simulated galaxies fed by cold streams}",
      journal = {\mnras},
         year = 2011,
        month = dec,
       volume = {418},
       number = {3},
        pages = {1796-1821},
          doi = {10.1111/j.1365-2966.2011.19599.x},
archivePrefix = {arXiv},
       eprint = {1103.2130},
 primaryClass = {astro-ph.CO},
       adsurl = {https://ui.adsabs.harvard.edu/abs/2011MNRAS.418.1796F}
}

@article{peroux_nature_2002,
       author = {{P{\'e}roux}, C{\'e}line and {Dessauges-Zavadsky}, Miroslava and {Kim}, TaeSun and {McMahon}, Richard G. and {D'Odorico}, Sandro},
        title = "{Nature and Properties of sub-DLAs (absorbers with {}10$^{19}$ {\ensuremath{\leq}} N(HI) {\ensuremath{\leq}} 2 * {}10$^{20}$ cm$^{-2}$)}",
      journal = {\apss},
         year = 2002,
        month = jul,
       volume = {281},
       number = {1},
        pages = {543-544},
          doi = {10.1023/A:1019591430144},
       adsurl = {https://ui.adsabs.harvard.edu/abs/2002Ap&SS.281..543P}
}

@article{wolfe_damped_2005,
       author = {{Wolfe}, Arthur M. and {Gawiser}, Eric and {Prochaska}, Jason X.},
        title = "{Damped Ly {\ensuremath{\alpha}} Systems}",
      journal = {\araa},
         year = 2005,
        month = sep,
       volume = {43},
       number = {1},
        pages = {861-918},
          doi = {10.1146/annurev.astro.42.053102.133950},
archivePrefix = {arXiv},
       eprint = {astro-ph/0509481},
 primaryClass = {astro-ph},
       adsurl = {https://ui.adsabs.harvard.edu/abs/2005ARA&A..43..861W}
}

@article{tytler_qso_1982,
       author = {{Tytler}, D.},
        title = "{QSO Lyman limit absorption}",
      journal = {\nat},
         year = 1982,
        month = jul,
       volume = {298},
       number = {5873},
        pages = {427-432},
          doi = {10.1038/298427a0},
       adsurl = {https://ui.adsabs.harvard.edu/abs/1982Natur.298..427T}
}

@article{lynds_absorption-line_1971,
       author = {{Lynds}, Roger},
        title = "{The Absorption-Line Spectrum of 4c 05.34}",
      journal = {\apjl},
         year = 1971,
        month = mar,
       volume = {164},
        pages = {L73},
          doi = {10.1086/180695},
       adsurl = {https://ui.adsabs.harvard.edu/abs/1971ApJ...164L..73L}
}

@ARTICLE{Vernet2011,
       author = {{Vernet}, J. and {Dekker}, H. and {D'Odorico}, S. and {Kaper}, L. and {Kjaergaard}, P. and {Hammer}, F. and {Randich}, S. and {Zerbi}, F. and {Groot}, P.~J. and {Hjorth}, J. and {Guinouard}, I. and {Navarro}, R. and {Adolfse}, T. and {Albers}, P.~W. and {Amans}, J. -P. and {Andersen}, J.~J. and {Andersen}, M.~I. and {Binetruy}, P. and {Bristow}, P. and {Castillo}, R. and {Chemla}, F. and {Christensen}, L. and {Conconi}, P. and {Conzelmann}, R. and {Dam}, J. and {de Caprio}, V. and {de Ugarte Postigo}, A. and {Delabre}, B. and {di Marcantonio}, P. and {Downing}, M. and {Elswijk}, E. and {Finger}, G. and {Fischer}, G. and {Flores}, H. and {Fran{\c{c}}ois}, P. and {Goldoni}, P. and {Guglielmi}, L. and {Haigron}, R. and {Hanenburg}, H. and {Hendriks}, I. and {Horrobin}, M. and {Horville}, D. and {Jessen}, N.~C. and {Kerber}, F. and {Kern}, L. and {Kiekebusch}, M. and {Kleszcz}, P. and {Klougart}, J. and {Kragt}, J. and {Larsen}, H.~H. and {Lizon}, J. -L. and {Lucuix}, C. and {Mainieri}, V. and {Manuputy}, R. and {Martayan}, C. and {Mason}, E. and {Mazzoleni}, R. and {Michaelsen}, N. and {Modigliani}, A. and {Moehler}, S. and {M{\o}ller}, P. and {Norup S{\o}rensen}, A. and {N{\o}rregaard}, P. and {P{\'e}roux}, C. and {Patat}, F. and {Pena}, E. and {Pragt}, J. and {Reinero}, C. and {Rigal}, F. and {Riva}, M. and {Roelfsema}, R. and {Royer}, F. and {Sacco}, G. and {Santin}, P. and {Schoenmaker}, T. and {Spano}, P. and {Sweers}, E. and {Ter Horst}, R. and {Tintori}, M. and {Tromp}, N. and {van Dael}, P. and {van der Vliet}, H. and {Venema}, L. and {Vidali}, M. and {Vinther}, J. and {Vola}, P. and {Winters}, R. and {Wistisen}, D. and {Wulterkens}, G. and {Zacchei}, A.},
        title = "{X-shooter, the new wide band intermediate resolution spectrograph at the ESO Very Large Telescope}",
      journal = {\aap},
         year = 2011,
        month = dec,
       volume = {536},
          eid = {A105},
        pages = {A105},
          doi = {10.1051/0004-6361/201117752},
archivePrefix = {arXiv},
       eprint = {1110.1944},
 primaryClass = {astro-ph.IM},
       adsurl = {https://ui.adsabs.harvard.edu/abs/2011A&A...536A.105V}
}

@ARTICLE{Lopez2016,
       author = {{L{\'o}pez}, S. and {D'Odorico}, V. and {Ellison}, S.~L. and {Becker}, G.~D. and {Christensen}, L. and {Cupani}, G. and {Denney}, K.~D. and {P{\^a}ris}, I. and {Worseck}, G. and {Berg}, T.~A.~M. and {Cristiani}, S. and {Dessauges-Zavadsky}, M. and {Haehnelt}, M. and {Hamann}, F. and {Hennawi}, J. and {Ir{\v{s}}i{\v{c}}}, V. and {Kim}, T. -S. and {L{\'o}pez}, P. and {Lund Saust}, R. and {M{\'e}nard}, B. and {Perrotta}, S. and {Prochaska}, J.~X. and {S{\'a}nchez-Ram{\'\i}rez}, R. and {Vestergaard}, M. and {Viel}, M. and {Wisotzki}, L.},
        title = "{XQ-100: A legacy survey of one hundred 3.5 {\ensuremath{\lesssim}} z {\ensuremath{\lesssim}} 4.5 quasars observed with VLT/X-shooter}",
      journal = {\aap},
         year = 2016,
        month = oct,
       volume = {594},
          eid = {A91},
        pages = {A91},
          doi = {10.1051/0004-6361/201628161},
archivePrefix = {arXiv},
       eprint = {1607.08776},
 primaryClass = {astro-ph.GA},
       adsurl = {https://ui.adsabs.harvard.edu/abs/2016A&A...594A..91L}
}

@ARTICLE{Schechter1976,
   author = {{Schechter}, P.},
    title = "{An Analytic Expression for the Luminosity Function for Galaxies}",
  journal = {\apj},
     year = 1976,
    month = nov,
   volume = 203,
    pages = {297},
      doi = {10.1086/154079},
   adsurl = {https://ui.adsabs.harvard.edu/abs/1976ApJ...203..297S}
}

@ARTICLE{Davis1983,
       author = {{Davis}, M. and {Peebles}, P.~J.~E.},
        title = "{A survey of galaxy redshifts. V. The two-point position and velocity correlations.}",
      journal = {\apj},
         year = 1983,
        month = apr,
       volume = {267},
        pages = {465-482},
          doi = {10.1086/160884},
       adsurl = {https://ui.adsabs.harvard.edu/abs/1983ApJ...267..465D}
}

@ARTICLE{Morrison2024,
       author = {{Morrison}, Sean and {Som}, Debopam and {Pieri}, Matthew M. and {P{\'e}rez-R{\`a}fols}, Ignasi and {Blomqvist}, Michael},
        title = "{A strong blend in the morning: studying the circumgalactic medium before cosmic noon with strong, blended Ly {\ensuremath{\alpha}} forest systems}",
      journal = {\mnras},
         year = 2024,
        month = jul,
       volume = {532},
       number = {1},
        pages = {32-59},
          doi = {10.1093/mnras/stae1418},
archivePrefix = {arXiv},
       eprint = {2309.06813},
 primaryClass = {astro-ph.CO},
       adsurl = {https://ui.adsabs.harvard.edu/abs/2024MNRAS.532...32M}
}

@ARTICLE{Perez2023,
       author = {{P{\'e}rez-R{\`a}fols}, Ignasi and {Pieri}, Matthew M. and {Blomqvist}, Michael and {Morrison}, Sean and {Som}, Debopam and {Cuceu}, Andrei},
        title = "{The cross-correlation of galaxies in absorption with the Lyman {\ensuremath{\alpha}} forest}",
      journal = {\mnras},
         year = 2023,
        month = sep,
       volume = {524},
       number = {1},
        pages = {1464-1477},
          doi = {10.1093/mnras/stad1994},
archivePrefix = {arXiv},
       eprint = {2210.02973},
 primaryClass = {astro-ph.CO},
       adsurl = {https://ui.adsabs.harvard.edu/abs/2023MNRAS.524.1464P}
}

@ARTICLE{Pieri2014,
       author = {{Pieri}, Matthew M. and {Mortonson}, Michael J. and {Frank}, Stephan and {Crighton}, Neil and {Weinberg}, David H. and {Lee}, Khee-Gan and {Noterdaeme}, Pasquier and {Bailey}, Stephen and {Busca}, Nicolas and {Ge}, Jian and {Kirkby}, David and {Lundgren}, Britt and {Mathur}, Smita and {P{\^a}ris}, Isabelle and {Palanque-Delabrouille}, Nathalie and {Petitjean}, Patrick and {Rich}, James and {Ross}, Nicholas P. and {Schneider}, Donald P. and {York}, Donald G.},
        title = "{Probing the circumgalactic medium at high-redshift using composite BOSS spectra of strong Lyman {\ensuremath{\alpha}} forest absorbers}",
      journal = {\mnras},
         year = 2014,
        month = jun,
       volume = {441},
       number = {2},
        pages = {1718-1740},
          doi = {10.1093/mnras/stu577},
archivePrefix = {arXiv},
       eprint = {1309.6768},
 primaryClass = {astro-ph.CO},
       adsurl = {https://ui.adsabs.harvard.edu/abs/2014MNRAS.441.1718P}
}

@ARTICLE{Becker2013,
       author = {{Becker}, George D. and {Hewett}, Paul C. and {Worseck}, G{\'a}bor and {Prochaska}, J. Xavier},
        title = "{A refined measurement of the mean transmitted flux in the Ly{\ensuremath{\alpha}} forest over 2 < z < 5 using composite quasar spectra}",
      journal = {\mnras},
         year = 2013,
        month = apr,
       volume = {430},
       number = {3},
        pages = {2067-2081},
          doi = {10.1093/mnras/stt031},
archivePrefix = {arXiv},
       eprint = {1208.2584},
 primaryClass = {astro-ph.CO},
       adsurl = {https://ui.adsabs.harvard.edu/abs/2013MNRAS.430.2067B}
}

@ARTICLE{Pieri2010,
       author = {{Pieri}, Matthew M. and {Frank}, Stephan and {Weinberg}, David H. and {Mathur}, Smita and {York}, Donald G.},
        title = "{The Composite Spectrum of Strong Ly{\ensuremath{\alpha}} Forest Absorbers}",
      journal = {\apjl},
         year = 2010,
        month = nov,
       volume = {724},
       number = {1},
        pages = {L69-L73},
          doi = {10.1088/2041-8205/724/1/L69},
archivePrefix = {arXiv},
       eprint = {1001.5282},
 primaryClass = {astro-ph.CO},
       adsurl = {https://ui.adsabs.harvard.edu/abs/2010ApJ...724L..69P}
}

@misc{santos_finding_2025,
    title = {Finding {Halos} in the {Lyman}a forest},
    url = {http://arxiv.org/abs/2507.08940},
    doi = {10.48550/arXiv.2507.08940},
    urldate = {2025-07-15},
    publisher = {arXiv},
    author = {Muñoz Santos, Duarte and Pieri, Matthew M. and Nelson, Dylan and Weng, Simon and Hu, Teng and Bernal, Manuel F. Ruiz-Herrera},
    month = jul,
    year = {2025},
    note = {arXiv:2507.08940 [astro-ph]},
}

@article{rudie_gaseous_2012,
    title = {The {Gaseous} {Environment} of {High}-z {Galaxies}: {Precision} {Measurements} of {Neutral} {Hydrogen} in the {Circumgalactic} {Medium} of z {\textasciitilde} 2-3 {Galaxies} in the {Keck} {Baryonic} {Structure} {Survey}},
    volume = {750},
    issn = {0004-637X},
    shorttitle = {The {Gaseous} {Environment} of {High}-z {Galaxies}},
    url = {https://ui.adsabs.harvard.edu/abs/2012ApJ...750...67R},
    doi = {10.1088/0004-637X/750/1/67},
    urldate = {2025-10-27},
    journal = {The Astrophysical Journal},
    author = {Rudie, Gwen C. and Steidel, Charles C. and Trainor, Ryan F. and Rakic, Olivera and Bogosavljević, Milan and Pettini, Max and Reddy, Naveen and Shapley, Alice E. and Erb, Dawn K. and Law, David R.},
    month = may,
    year = {2012},
    note = {Publisher: IOP
ADS Bibcode: 2012ApJ...750...67R},
    pages = {67},
}

@article{rakic_neutral_2012,
    title = {Neutral {Hydrogen} {Optical} {Depth} near {Star}-forming {Galaxies} at z ≈ 2.4 in the {Keck} {Baryonic} {Structure} {Survey}},
    volume = {751},
    issn = {0004-637X},
    url = {https://ui.adsabs.harvard.edu/abs/2012ApJ...751...94R},
    doi = {10.1088/0004-637X/751/2/94},
    urldate = {2025-10-27},
    journal = {The Astrophysical Journal},
    author = {Rakic, Olivera and Schaye, Joop and Steidel, Charles C. and Rudie, Gwen C.},
    month = jun,
    year = {2012},
    note = {Publisher: IOP
ADS Bibcode: 2012ApJ...751...94R},
    pages = {94},
}

@article{bielby_vlt_2013,
    title = {The {VLT} {LBG} {Redshift} {Survey} - {III}. {The} clustering and dynamics of {Lyman}-break galaxies at z ∼ 3},
    volume = {430},
    issn = {0035-8711},
    url = {https://ui.adsabs.harvard.edu/abs/2013MNRAS.430..425B},
    doi = {10.1093/mnras/sts639},
    urldate = {2025-10-27},
    journal = {Monthly Notices of the Royal Astronomical Society},
    author = {Bielby, R. and Hill, M. D. and Shanks, T. and Crighton, N. H. M. and Infante, L. and Bornancini, C. G. and Francke, H. and Héraudeau, P. and Lambas, D. G. and Metcalfe, N. and Minniti, D. and Padilla, N. and Theuns, T. and Tummuangpak, P. and Weilbacher, P.},
    month = mar,
    year = {2013},
    note = {Publisher: OUP
ADS Bibcode: 2013MNRAS.430..425B},
    pages = {425--449},
}

@article{rakic_measurement_2013,
    title = {A measurement of galaxy halo mass from the surrounding {H} i {Lyα} absorption},
    volume = {433},
    issn = {0035-8711},
    url = {https://doi.org/10.1093/mnras/stt950},
    doi = {10.1093/mnras/stt950},
    number = {4},
    urldate = {2025-10-27},
    journal = {Monthly Notices of the Royal Astronomical Society},
    author = {Rakic, Olivera and Schaye, Joop and Steidel, Charles C. and Booth, C. M. and Dalla Vecchia, Claudio and Rudie, Gwen C.},
    month = aug,
    year = {2013},
    pages = {3103--3114},
}

@article{herrero_alonso_clustering_2023,
    title = {Clustering dependence on {Lyα} luminosity from {MUSE} surveys at 3 {\textless} z {\textless} 6},
    volume = {671},
    issn = {0004-6361},
    url = {https://ui.adsabs.harvard.edu/abs/2023A&A...671A...5H},
    doi = {10.1051/0004-6361/202244693},
    urldate = {2025-10-27},
    journal = {Astronomy and Astrophysics},
    author = {Herrero Alonso, Y. and Miyaji, T. and Wisotzki, L. and Krumpe, M. and Matthee, J. and Schaye, J. and Aceves, H. and Kusakabe, H. and Urrutia, T.},
    month = mar,
    year = {2023},
    note = {Publisher: EDP
ADS Bibcode: 2023A\&A...671A...5H},
    pages = {A5},
}

@article{muzahid_musequbes_2021,
    title = {{MUSEQuBES}: characterizing the circumgalactic medium of redshift ≈3.3 {Ly} α emitters},
    volume = {508},
    issn = {0035-8711},
    shorttitle = {{MUSEQuBES}},
    url = {https://ui.adsabs.harvard.edu/abs/2021MNRAS.508.5612M},
    doi = {10.1093/mnras/stab2933},
    urldate = {2025-10-30},
    journal = {Monthly Notices of the Royal Astronomical Society},
    author = {Muzahid, Sowgat and Schaye, Joop and Cantalupo, Sebastiano and Marino, Raffaella Anna and Bouché, Nicolas F. and Johnson, Sean and Maseda, Michael and Wendt, Martin and Wisotzki, Lutz and Zabl, Johannes},
    month = dec,
    year = {2021},
    note = {Publisher: OUP
ADS Bibcode: 2021MNRAS.508.5612M},
    pages = {5612--5637},
}

@article{Herrera2025,
	author = {{Herrera}, Danisbel and {Gawiser}, Eric and {Benda}, Barbara and {Firestone}, Nicole M. and {Ramakrishnan}, Vandana and {Moon}, Byeongha and {Lee}, Kyoung-Soo and {Park}, Changbom and {Valdes}, Francisco and {Yang}, Yujin and {Artale}, Mar{\'\i}a Celeste and {Ciardullo}, Robin and {Gronwall}, Caryl and {Guaita}, Lucia and {Hwang}, Ho Seong and {Kennedy}, Jacob and {Kumar}, Ankit and {Zabludoff}, Ann},
        title = "{ODIN: Clustering Analysis of 14,000 Ly{\ensuremath{\alpha}}-emitting Galaxies at z = 2.4, 3.1, and 4.5}",
      journal = {\apjl},
         year = 2025,
        month = aug,
       volume = {988},
       number = {2},
          eid = {L57},
        pages = {L57},
          doi = {10.3847/2041-8213/adec82},
 primaryClass = {astro-ph.GA},
       adsurl = {https://ui.adsabs.harvard.edu/abs/2025ApJ...988L..57H}
}

@ARTICLE{Planck2016,
       author = {{Planck Collaboration} and {Adam}, R. and {Ade}, P.~A.~R. and {et al.}},
        title = "{Planck 2015 results. I. Overview of products and scientific results}",
      journal = {\aap},
         year = 2016,
        month = may,
       volume = {594},
          eid = {A1},
        pages = {A1},
          doi = {10.1051/0004-6361/201527101},
archivePrefix = "arXiv",
       eprint = {1502.01582},
 primaryClass = "astro-ph.CO",
       adsurl = {https://ui.adsabs.harvard.edu/abs/2016A&A...594A...1P}
}

@ARTICLE{Matthee+2024,
       author = {{Matthee}, Jorryt and {Golling}, Christopher and {Mackenzie}, Ruari and {Pezzulli}, Gabriele and {Lilly}, Simon and {Schaye}, Joop and {Bacon}, Roland and {Kusakabe}, Haruka and {Urrutia}, Tanya and {Boogaard}, Leindert and {Brinchmann}, Jarle and {Maseda}, Michael V. and {Garel}, Thibault and {Bouch{\'e}}, Nicolas F. and {Wisotzki}, Lutz},
        title = "{Large-scale excess H I absorption around z {\ensuremath{\approx}} 4 galaxies detected in a background galaxy spectrum in the MUSE eXtremely deep field}",
      journal = {\mnras},
         year = 2024,
        month = apr,
       volume = {529},
       number = {3},
        pages = {2794-2806},
          doi = {10.1093/mnras/stae673},
archivePrefix = {arXiv},
       eprint = {2305.15346},
 primaryClass = {astro-ph.GA},
       adsurl = {https://ui.adsabs.harvard.edu/abs/2024MNRAS.529.2794M}
}

\end{document}